\documentclass[preprint,12pt]{elsarticle}
\usepackage[greek,english]{babel}
\usepackage[LGR,T1]{fontenc}
\usepackage[latin1]{inputenc}
\usepackage{amssymb}
\usepackage{amsmath}
\usepackage{graphicx,xcolor}
\usepackage{amsthm}
\usepackage{mathrsfs}
\usepackage{array}
\usepackage{subfigure}
\usepackage{pdfpages}
\usepackage{url}
\usepackage{color}
\usepackage[ruled]{algorithm2e}
\usepackage{natbib}
\usepackage{setspace}
\usepackage{geometry}
\newtheorem{theorem}{Theorem}
\newtheorem{lemma}{Lemma}
\newtheorem{proposition}{Proposition}
\newtheorem{definition}{Definition}
\begin{document}

\begin{frontmatter}

%=======================================================================================
\title{Recursive simplex stars}
%=======================================================================================

\author{Guillaume Deffuant}
\address{Irstea - LISC, 9 avenue Blaise Pascal 63178 Aubi\`{e}re, France}
%\email{guillaume.deffuant@irstea.fr}

\begin{abstract}
This paper proposes a new method which builds a simplex based approximation of a $d-1$-dimensional manifold $M$ separating a $d$-dimensional compact set into two parts, and an efficient algorithm classifying points according to this approximation. In a first variant, the approximation is made of simplices that are defined in the cubes of a regular grid covering the compact set, from boundary points that approximate the intersection between $M$ and the edges of the cubes. All the simplices defined in a cube share the barycentre of the boundary points located in the cube and include simplices similarly defined in cube facets, and so on recursively. In a second variant, the Kuhn triangulation is used to break the cubes into simplices and the approximation is defined in these simplices from the boundary points computed on their edges, with the same principle. Both the approximation in cubes and in simplices define a separating surface on the whole grid and classifying a point on one side or the other of this surface requires only a small number (at most $d$) of simple tests. Under some conditions on the definition of the boundary points and on the reach of the surface to approximate, for both variants the Hausdorff distance between $M$ and its approximation decreases like $\mathcal{O}(d n_G^{-2})$, where $n_G$ is the number of points on each axis of the grid. The approximation in cubes requires computing less boundary points than the approximation in simplices but the latter is always a manifold and is more accurate for a given value of $n_G$. The paper reports tests of the method when varying $n_G$ and the dimensionality of the space (up to 9). 
\end{abstract}

\begin{keyword}
classification \sep marching cubes \sep simplex star
\end{keyword}

\maketitle

\end{frontmatter}

\SetAlgoSkip{bigskip}
%\tableofcontents
%
%--------------------------------------------------------------------------------------------------------
\section{Introduction.}
%--------------------------------------------------------------------------------------------------------
This paper addresses the following problems: 
\begin{itemize}
	\item How to approximate a $d-1$-dimensional manifold $M$ separating a $d$-dimensional compact set $X$ into two parts (one labelled -1, the other +1), using as efficiently as possible an oracle which, to any point of $X$, provides the value stating on which part of $X$ the point is located ? 
	\item How to define an efficient algorithm computing the classification of a point by the approximate separation ?
\end{itemize}

The main motivation is to improve algorithms derived from Viability Theory \cite{Aubin1991,Aubin2011}. This theory, which provides methods and tools for maintaining a dynamical system within a constraint set, is used in many fields such as sustainability management \cite{Martin2004,Delara2008,Deffuant2011,Mathias2015,Heitzig2016,Oubraham2018}, economics \cite{Aubin1997} or food processing \cite{Mesmoudi2014}. The main algorithms derived from Viability Theory \citep{Saint-Pierre1994,Deffuant2007} iterate the computation of approximate classification functions using the vertices of a grid labelled into two classes. The approximate classification methods currently used are the nearest vertex of the grid \cite{Saint-Pierre1994} or machine learning techniques such as support vector machines \cite{Deffuant2007} or $k$-$d$ trees \cite{Alvarez2016}. Our main purpose is to develop a more efficient method. 

Nevertheless, deriving efficient approximate classification can be useful in other contexts. For instance, when a classification requires a heavy or difficult process, it is often interesting to compute an approximate but lighter classification function, based on a limited set of well chosen classified points. This problem is a particular case of meta (or surrogate) modelling. The field of reliability in material sciences for instance develops specific techniques to build such meta-models \cite{Lemaire2009}. 

A problem closely related to the approximation of a classification boundary is the approximation of an isosurface, defined as the set of points such that $f(x)=0$, $f$ being a continuous function from the considered space into $\mathbb{R}$. In this problem, it is possible to define local linear approximations of $f$ around the values $f(v)$ of the vertices $v$ of the grid, which is not possible in the approximation of a classification boundary because the values at the vertices are either -1 or +1.

In 3 dimensions, the problem of approximating an isosurface is very common, for instance to visualise surfaces from scanners or magnetic resonance imaging measurements, and several techniques are available. In particular, the algorithms deriving from the marching cubes \cite{Lorensen1987} (see \cite{newman2006} for a review) build simplex-based surfaces. They firstly compute the boundary points approximating the intersections between the isosurface and the edges of a regular grid, generally using a linear interpolation. Then the marching cubes generally use a table of rules specifying the connections between boundary points to define a simplex based separating surface in each cube configuration. Once solved the problems of consistency between cubes \cite{Nielson1991}, these techniques represent efficiently the surface of 3D objects. Some variants include an adaptive refinement of the grid in order to guarantee that the approximation is isotopic with the surface to approximate \cite{Plantinga2004}. 

However, extending these methods to spaces of more than 3 dimensions faces serious difficulties as the number of cube configurations is in $2^{2^d}$, which leads to a very high number of rules specifying the simplices by cube configuration. For instance in 6 dimensions, there are $2^{64}$ cube configurations which is beyond any current computer storage capacities. Moreover, the number of simplices grows exponentially with the dimensionality and so does the necessary memory space to store them. Currently, as far as we know, the available methods of marching cubes in arbitrary dimensionality are:
\begin{itemize}
	\item Breaking cubes into simplices \cite{Weigle1996,Weigle1998,Min2003}. This addresses the problem of the fast growth of the table of rules mentioned earlier, because the number of configurations in a simplex is much lower (it varies as $\frac{d+1}{2}$) than in a cube. However, a $d$-dimensional cube breaks into between $d$! or $2^{d-1}d$! simplices, depending on the decomposition used \cite{Weigle1998}. The cited papers show examples in at most 4 dimensions.
	\item Defining the simplices in a cube from the convex hull of a set of points including the boundary points and some cube vertices \cite{Lachaud2000,Bhaniramka2004}. The simplices are defined with a single rule but computing the convex hull and storing the corresponding simplices is computationally demanding when $d$ increases. Again, \cite{Bhaniramka2004} shows only examples up to 4 dimensions.
\end{itemize}

A different approach builds on the principles of Delaunay triangulation and defines simplex based surfaces approximating a manifold without using a grid, from a sampling of points on this manifold. In addition to practical algorithms, the researchers studied the topological and geometric closeness between the approximation and the manifold \cite{Dey2008,Boissonnat2004,Chew2007}. Moreover, some variants are based on iterative sampling \cite{Chew1993,Boissonnat2005} adapting the density of the sampling to the local complexity of the shape. Recent variants of the approach \cite{Boissonnat2009,Boissonnat2011} approximate smooth manifolds of any dimensionality. However, the time complexity of the algorithm is exponential in $d'^2$ where $d'$ is the dimensionality of the manifold to approximate, which makes it difficult to apply practically even for moderate values of $d'$ (say $d' > 5$). The memory size needed to store the set of simplices also grows significantly with the dimensionality.

Moreover, none of these approaches considers the problem of using these approximations for a classification purpose. Indeed, when the number of simplices is very large, as expected when the dimensionality increases, computing the classification is also expected to become very demanding. Solving viability problems requires classifying large numbers of points, hence the efficiency of this procedure is crucial in this context. A specificity of this paper is precisely to propose an efficient classification algorithm adapted to specific structures of simplices.

The method proposed in this paper uses a regular grid like the marching cubes and generalises the method of centroids \cite{Ning1993} to an arbitrary dimensionality. Like the dual marching cubes \cite{Ashida2003,Nielson2004,Gress2004} it adds new points in cubes and faces. A noticeable difference with the standard marching cubes is that the boundary points cannot be approximated linearly and successive dichotomies are used instead. %The main principle is to define recursive simplex stars (resistars) in the grid cubes. 

In a first variant of the proposed method, the simplices are defined in cubes of the grid, from the boundary points located on their edges. All simplices share the barycentre of the cube boundary points as a common vertex (thus shaping a "simplex star") and include simplices of lower dimensionality similarly defined in cube facets. The recursion ends when the considered face is an edge of the cube containing a boundary point. 

This method can be related to the barycentric subdivision which divides an arbitrary convex polytope into simplices sharing the barycentre of the polytope's vertices and this operation can be recursively applied to the faces of the polytope. The main difference is that the simplices of the proposed method are defined in the faces of the cube in which the boundary points are located, not in the faces of the polytope that they define.

In the proposed approach, there is only one rule deriving the simplices in a cube (or a face) whatever the dimensionality. The simplices can easily be enumerated, going through all the faces of a cube. However, when there are 2D faces including 4 boundary points, this method "glues" together surfaces that should remain separated. This creates a non-manifold approximation which should be avoided in many applications. %This problem also occurs in the marching cube dual contouring approach \cite{Nielson2004}.

The second variant of the method addresses this problem. It uses the Kuhn triangulation to break the cubes into simplices like in \cite{Weigle1998}. The boundary points are computed on the edges of these simplices and the approximation is defined as previously from these boundary points, using the faces of a simplex instead of the faces of a cube. This variant always defines a manifold.

The paper underlines the following properties of both variants:
\begin{itemize}
\item It is possible to compute the classification of a point with at most $d$ relatively simple operations; 

\item Under some conditions on the computation of the boundary points and on the smoothness of $M$, the Hausdorff distance between $M$ and its approximation decreases like $\mathcal{O}(dn_G^{-2})$, $n_G$ being the number of points on each axis of the grid.  

\end{itemize}

The remaining of the paper is organised as follows: Section 2 presents the variant of the approximation defined in cubes of the grid and its classification algorithm, section 3 does the same for the variant defined in the Kuhn simplices of the grid, section 4 establishes the theorem about the approximation accuracy, section 5 reports the results of a series of tests of the method when varying the space dimensionality and the size of the grid and finally section 6 discusses these results and potential extensions.

%=======================================================
\section{Manifold approximation with resistars in cubes.}
%=======================================================

Let $M$ be a $d-1$-dimensional manifold separating the compact set $X = [0, 1]^d$ into two parts (one labelled $-1$ the other $+1$) and let $\mathcal{M}: X \rightarrow \left\{-1, 0, +1\right\}$, the function which outputs 0 if $x$ belongs to $M$ and otherwise the label of the part of $X$ in which $x$ is located.  

We consider a regular grid $G$ of $n_G^d$ points covering $X$ and its boundary. The distance between two adjacent points of the grid is $\epsilon = \frac{1}{n_G-1}$. The cubes of the grid are $d$-dimensional cubes of edge size $\epsilon$ whose vertices are points of the grid. The faces of the grid are the faces of these cubes (cubic polytopes of dimensionality lower than $d$, of edge size $\epsilon$ and whose vertices are points of the grid).

 It is supposed that none of the grid points belongs to $M$. The function $\mathcal{M}$ is slightly modified if necessary in order to ensure this, as it is done in the marching cube approach.

The following notations are frequently used:
\begin{itemize}
\item The set of the $i$-dimensional faces of a grid cube or a face $C$ is denoted $\mathcal{F}_i(C)$;
\item For any polytope $P$, $\mathcal{V}(P)$ denotes the vertices of $P$; 
\item For a set $S$ of points of $X$, $\bar{S}$ denotes the barycentre of $S$, $[S]$ the convex hull of $S$ and $\partial S$ the boundary of $S$; 
\item For two sets $A$ and $B$ such that $B \subset A$, $A - B$ is the complementary of $B$ in A.
\end{itemize}

The next subsection focuses on the approximation in a single cube of the grid and the following subsection is devoted to the approximation on the whole grid.

%---------------------------------------------------------------
\subsection{c-resistar in a single cube.}
%---------------------------------------------------------------

\subsubsection{Boundary points.}
%---------------------------------------------------------------

The method requires first to define the boundary points in the cube.

\begin{definition}
Let $C$ be a $d$-dimensional cube of the grid and $n_e \geq 1$ be an integer. A boundary point $b_M([v_-,v_+])$ is defined on an edge $[v_-,v_+]$ of $C$ such that $\mathcal{M}(v_-) = -1$ and $\mathcal{M}(v_+)=+1$, as follows:
\begin{align}
 b_M([v_-,v_+]) = v_- + \frac{i+0.5}{n_e}(v_+ - v_-), 
\end{align}
Where $i \in \{0,..,n_e-1\}$ is such that:
\begin{align}
\mathcal{M}\left(v_- + \frac{i}{n_e}(v_+ - v_-)\right) = -1, \mbox{ and }
\mathcal{M}\left(v_- + \frac{i+1}{n_e}(v_+ - v_-)\right) = +1.
\end{align}
\end{definition}

In practice, the boundary points are determined by $q$ successive dichotomies with algorithm \ref{Algo:boundaryPoint}, and $n_e = 2^q$. Because $\mathcal{M}(v_-) = -1$ and $\mathcal{M}(v_+)=+1$, $M$ cuts the edge $[v_-,v_+]$ at least once, therefore $b_M([v_-,v_+])$ exists and there exists $b \in M \cap [v_-,v_+]$ such that:
\begin{align}
 \left\|b - b_M([v_-,v_+])\right\| \leq \frac{\left\|v_+ - v_-\right\|}{2 n_e} = 2^{-q -1}\epsilon.
\label{eq:precision}
\end{align}

If the manifold $M$ cuts the edge an odd number of times, this algorithm returns a boundary point which is close to one of the intersection points. Of course, if there is an even number of intersections, no boundary point is computed because the classification of the vertices by $\mathcal{M}$ is the same. As shown in section 4, under some conditions, it is possible to guarantee the accuracy of the approximation, despite the possibility of these situations.

\begin{algorithm}%[T]
%\dontprintsemicolon
\KwIn{$v_-$ , $v_+$, such that $[v_-,v_+]$ is an edge of the grid such that $\mathcal{M}(v_+) =1$ and $\mathcal{M}(v_-)= -1$, $q$ number of dichotomies}
	$p_+ \leftarrow v_+$; $p_- \leftarrow v_-$;\\
	\For{$ i \in \{1,.., q \}$}{
	$b_i \leftarrow \frac{p_- + p_+}{2}$;\\
	\eIf{$\mathcal{M}(b_i) = +1$}{$p_+ \leftarrow b_i$}
	{$p_- \leftarrow b_i$}
	}
\Return $\frac{p_- + p_+}{2}$;
\label{Algo:boundaryPoint}
\caption{Computation of boundary point $b$ located on edge $[v_-,v_+]$ with $q$ successive dichotomies.}
\end{algorithm}

For any face or cube $F$, we denote $B_M(F)$ the set of boundary points defined on the edges of $F$. 
For a boundary point $b \in [v_-, v_+]$, such that $\mathcal{M}(v_+) = +1$ and $\mathcal{M}(v_-) = -1$, $v_+$ is denoted $v_+(b)$ (resp. $v_-$ is denoted $v_-(b)$) and called the positive (resp. negative) vertex of $b$.

%The c-resistar is the $d-1$-dimensional surface defined as follows.
\subsubsection{Definition and main properties of c-resistars.}
%---------------------------------------------------------------

\begin{definition}
Let $C$ be a $d$-dimensional cube of the grid such that $B = B_M(C) \neq \emptyset$. The c-resistar approximation of $M$ in $C$, denoted $[B]^\star$, is the following set of simplices:
	
%\begin{align}
	%[B]^\star = \left\{ \textbf{[}b, (\bar{B}(F_i))_{i \in \{1,..,d\}}, \bar{B} \textbf{]},
	%[B]^\star = \left\{ [ b, (\bar{B}(F_i))_{2 \leq i \leq d-1}, \bar{B} ],
	%\begin{cases}
	%b \in B, b \in F_2,\\
	%F_i \in \mathcal{F}_i(C), F_{i} \subset F_{i+1}, 2 \leq i \leq d-1,
	%\end{cases}
	%\right\},
%\end{align}
\begin{align}
	[B]^\star =  \bigcup_{\{F_1,...,F_{d-1} \} \in \mathcal{F}^\star(B,C)}[\{\bar{B}_M(F_1),...,\bar{B}_M(F_{d-1}\} \cup \{ \bar{B} \} ],
\end{align}
with:
\begin{align}
\{F_1,...,F_{d-1} \} \in \mathcal{F}^\star(B,C) \Leftrightarrow
	\begin{cases}
	F_1 \in \mathcal{F}_1(C),	B_M(F_1) \neq \emptyset,\\
	F_i \in \mathcal{F}_i(C), F_{i-1} \subset F_{i}, 2 \leq i \leq d-1.
	\end{cases},
\end{align}
where $i$ is an integer.
\label{def:c-resistar}
\end{definition}

\begin{figure}
 	\centering
 	\begin{tabular}{cc}
 		\includegraphics[width=6 cm]{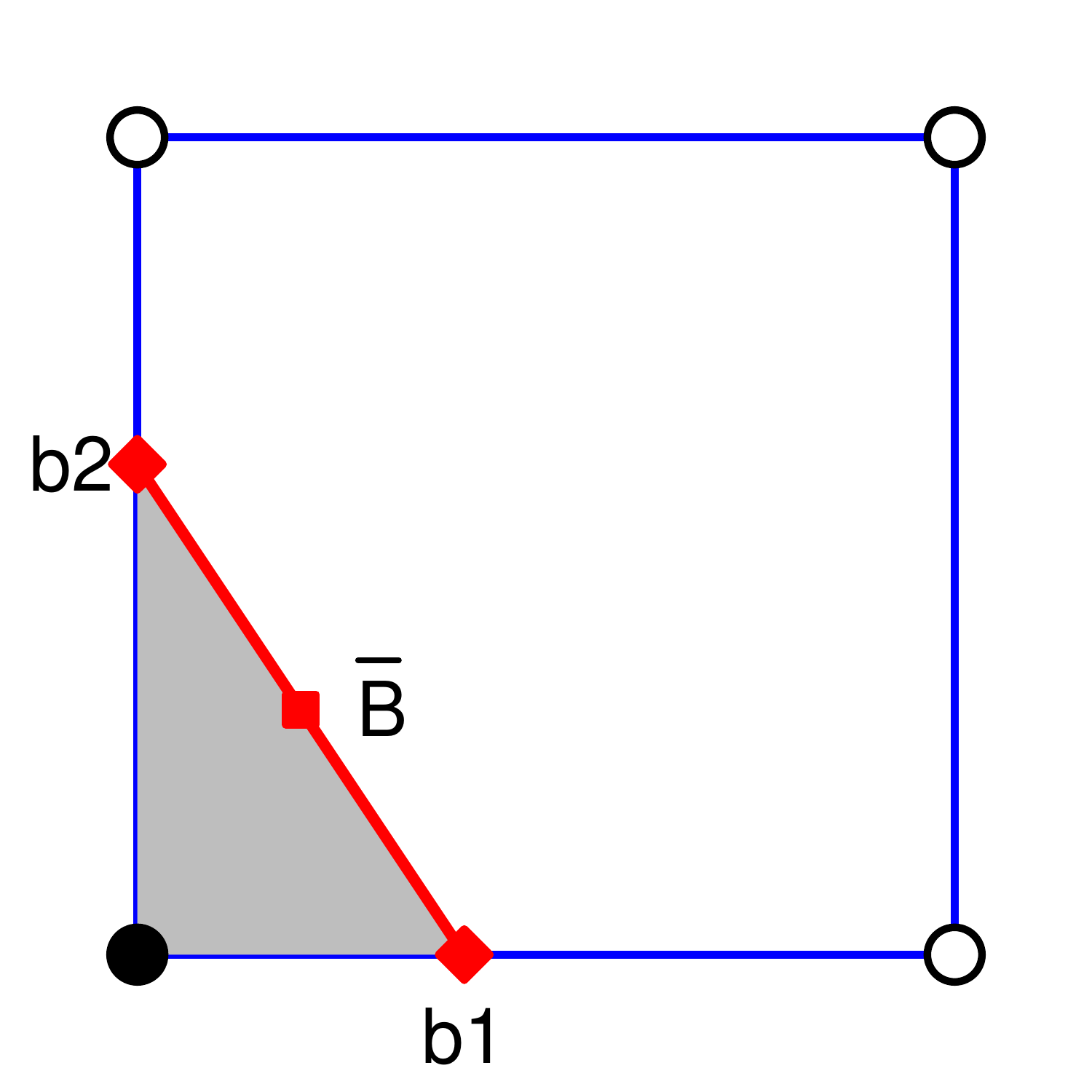} & 
 		\includegraphics[width= 6 cm]{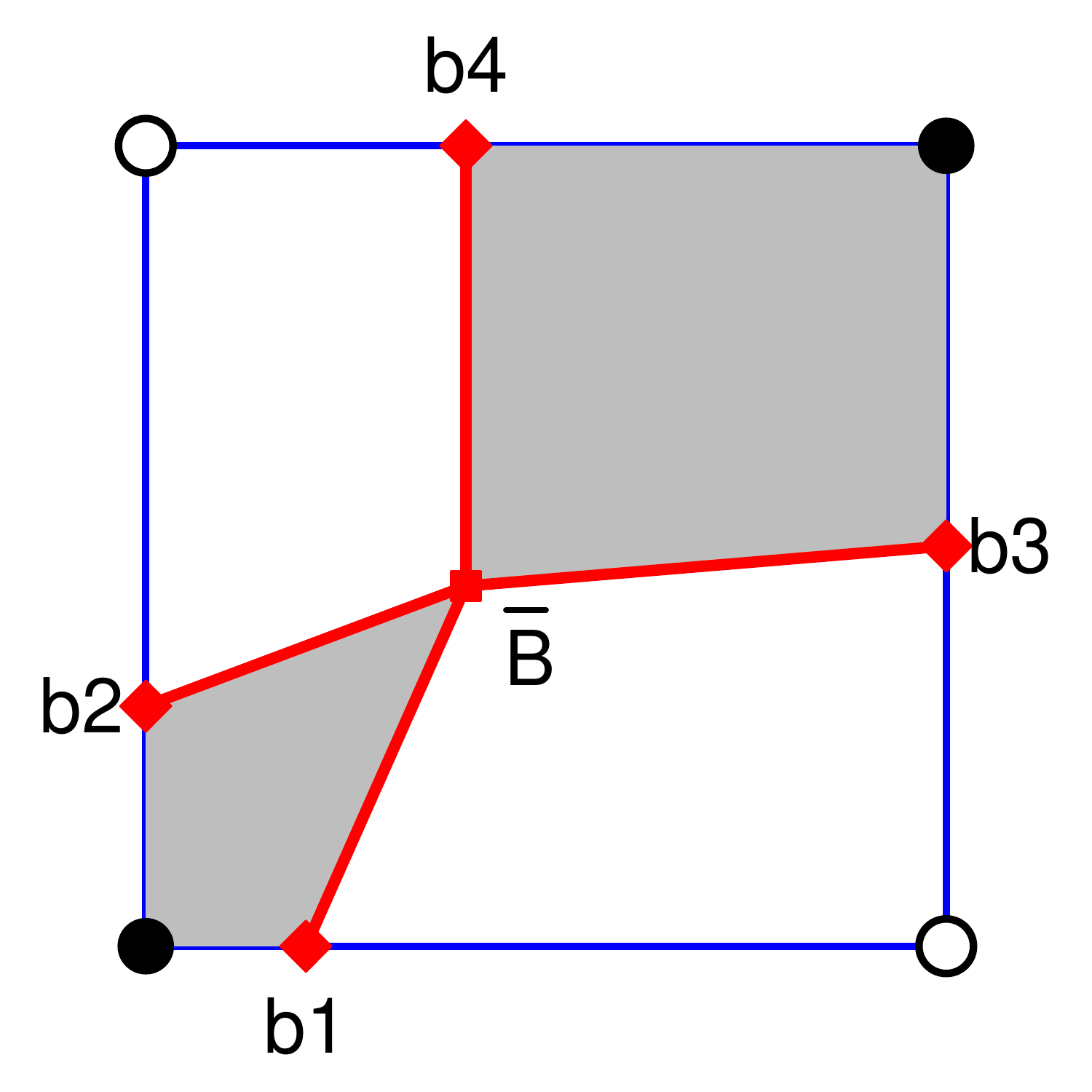} \\
		(a) & (b)
 	\end{tabular}
 	\caption{Examples of c-resistar in 2 D cubes. The cube vertices $v$ in white are such that $\mathcal{M}(v) = +1$ and those in black such that  $\mathcal{M}(v) = -1$. Panel (a): $[B]^\star = [b_1, \bar{B}]$ $\cup$  $ [b_2,\bar{B}] $. Panel (b): $[B]^\star =  [b_1,\bar{B}]$  $\cup$ $ [b_2, \bar{B}]$  $\cup$  $[b_3, \bar{B}]$  $\cup$ $[b_4, \bar{B}] $.}
 	\label{fig:cresistar2D}
 \end{figure} 

\begin{figure}
 	\centering
 		\includegraphics[width=9 cm]{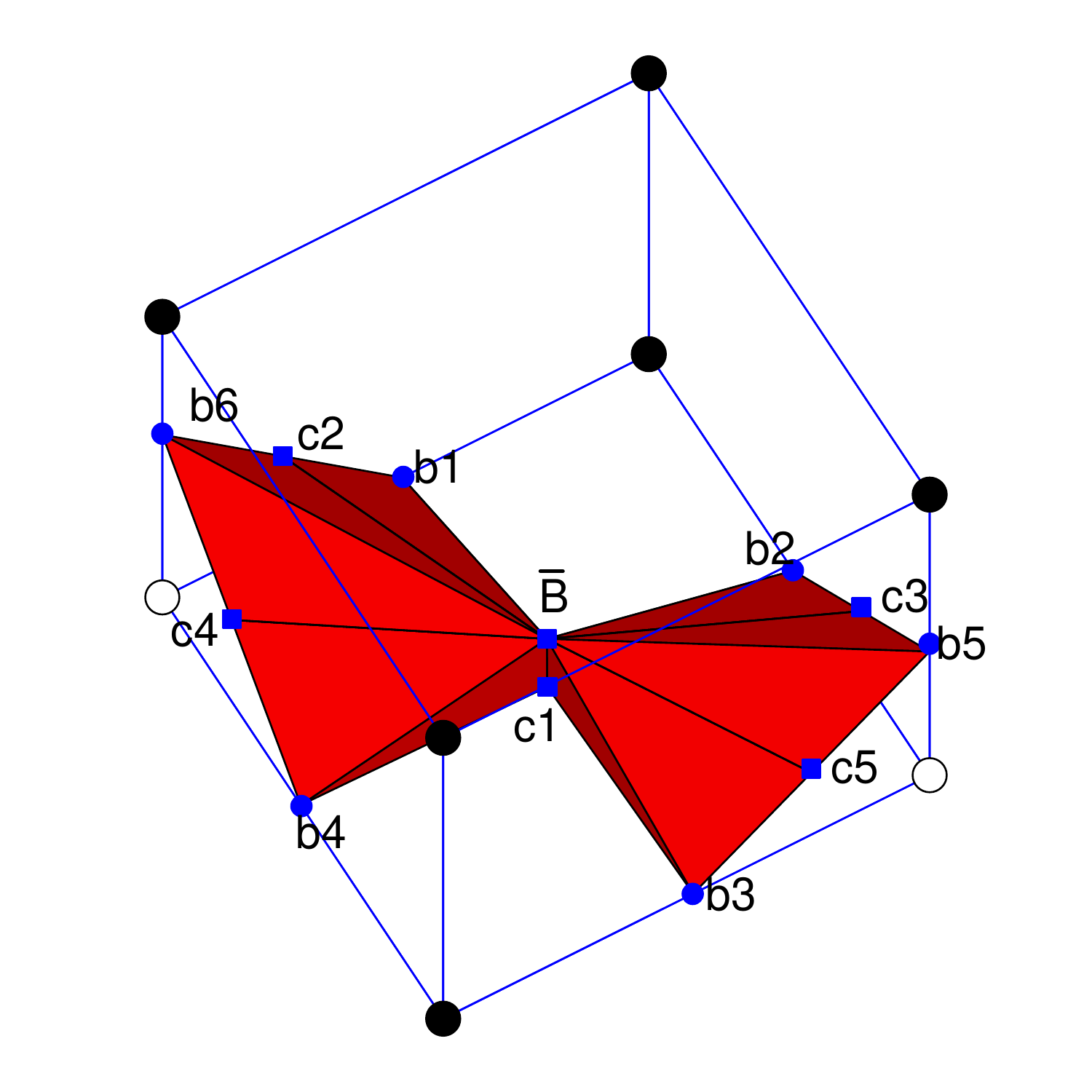} 
 	\caption{Example of a c-resistar in a 3 D cube, which includes 12 simplices. The vertices of these simplices are among the 6 boundary points $B = \{b_1,b_2, b_3, b_4, b_5, b_6\}$, the 5 barycentres of boundary points in 2D faces $\{c_1, c_2, c_3, c_4, c_5 \}$ and $\bar{B}$. $[B]^\star = [b_1, c_{1},\bar{B}]$ $\cup$ $[b_1, c_{2},\bar{B}]$  $\cup$  $[b_2, c_{1}, \bar{B}]$  $\cup$  $[b_2, c_{3}, \bar{B}]$  $\cup$  $[b_3,  c_{1}, \bar{B}]$ $\cup$ $[b_3,  c_{5}, \bar{B}]$  $\cup$  $[b_4, c_{1}, \bar{B}]$ $\cup$ $[b_4, c_{4}, \bar{B}]$ $\cup$ $[b_5, c_{3}, \bar{B}]$ $\cup$$[b_5, c_{5}, \bar{B}]$  $\cup$  $[b_6, c_{2}, \bar{B}]$ $\cup$  $[b_6, c_{4}, \bar{B}]$. The cube vertices $v$ in white are such that $\mathcal{M}(v) = +1$ and those in black such that  $\mathcal{M}(v) = -1$.}
 	\label{fig:cresistar3D}
 \end{figure} 

The word resistar stands for "recursive simplex star". Indeed, the barycentre $\bar{ B}$ of the boundary points is a vertex common to all the simplices which organises them as a star. This star is recursive because the common vertex $\bar{ B} $ is connected to simplices of lower dimensionality in the facets (faces of dimensionality $d-1$) of $C$ that share the barycentre $\bar{B}_M(F_{d-1})$ of the boundary points located in this facet, and so on recursively until reaching an edge $F_1$ of $C$ which include a boundary point (see examples on Figures \ref{fig:cresistar2D} and \ref{fig:cresistar3D}). We use the denomination c-resistar, with the prefix "c" standing for cube, in order to distinguish these resistars from the ones which are defined in Kuhn simplices, presented in section 3. 

Propositions \ref{prop:simplex} and \ref{prop:border} establish that the c-resistar is a $d-1$-dimensional surface without boundary inside the cube $C$.

\begin{proposition}
For all $\{ F_1, ..., F_{d-1} \} \in \mathcal{F}^\star(B,C)$, the set $[\{ \bar{B}_M(F_1),..., \bar{B}_M(F_{d-1}) \} \cup \{ \bar{B} \}]$ is a $d-1$-dimensional simplex.
\label{prop:simplex}
\end{proposition}

\textbf{Proof.}
Consider  $\{ F_1, ..., F_{d-1} \} \in \mathcal{F}^\star(B,C)$. Setting $F_d = C$, we will show that, for $1 \leq i \leq d-1$:
\begin{align}
\exists b_{i+1} \in B_M(F_{i+1}), b_{i+1} \notin B_M(F_i).
\end{align}

Let $F'_i \in \mathcal{F}_i(C)$ be the face opposite to $F_i$ in $F_{i+1}$. $F'_i$ is such that $\mathcal{V}(F_i) \cup \mathcal{V}(F'_i) = \mathcal{V}(F_{i+1})$ ($\mathcal{V}(F)$ being the set of vertices of face $F$). Two cases occur:
\begin{itemize}
	\item There are edges $[v,v']$ of $F'_i$ such that $\mathcal{M}(v).\mathcal{M}(v') = -1$, then, for each of these edges, there is a boundary point of $F_{i+1}$ which is not in $F_i$;
	\item All the vertices of $F'_i$ have the same classification by $\mathcal{M}$. By hypothesis, $B_M(F_i) \neq \emptyset$, thus there are vertices classified differently by $\mathcal{M}$ in $F_i$, hence there are vertices $v$ such that $\mathcal{M}(v) \neq \mathcal{M}(v')$, for  $v'$ vertex of $F'_i$. There are such couples $(v, v')$, for which $[v,v']$ is an edge of $F_{i+1}$, thus there is a boundary point $b_{i+1} \in [v,v']$ such that $b_{i+1} \in B_M(F_{i+1})$ and $b_{i+1} \notin F_i$.  
\end{itemize}
Therefore, for $1 \leq i \leq d-1$, $\bar{B}_M(F_{i+1}) \notin F_i$. Let $S = \{ \bar{B}_M(F_1),..., \bar{B}_M(F_{d-1}) \} \cup \{ \bar{B} \}$. The set $S$ includes $d$ affinely independent points, therefore $[S]$ is a $d-1$-dimensional simplex.
$\square$ %\vspace{0.5 cm}

\begin{proposition}
The boundary of the c-resistar $[B]^\star$ is included in the boundary of $C$.
\label{prop:border}
\end{proposition}

\textbf{Proof.}
Let $S = \{ \bar{B}_M(F_1),..., \bar{B}_M(F_{d-1}) \} \cup \{ \bar{B} \}$, with $\{ F_1, ..., F_{d-1} \} \in \mathcal{F}^\star(B,C)$. $[S]$ is a simplex of $[B]^\star$. Consider a facet $[S_F]$ of this simplex, with $S_F = S - \{ p \}$, $p \in S$. The following cases arise:
\begin{itemize}
	\item $p = \bar{B}$. By definition of $[B]^\star$, $S_F$ is included in the facet $F_{d-1}$ of $C$, thus $[S_F]$ is included in the boundary of $C$;
	\item $p = \bar{B}_M(F_i)$, $2 \leq i \leq d-1$. Assume that vectors $\{u_1,...,u_{i-1}\}$ are a basis $F_{i-1}$, vectors $\{u_1,...,u_{i-1}, u_i\}$ a basis of $F_{i}$ and vectors $\{u_1,...,u_{i-1}, u_i, u_{i+1}\}$ a basis of $F_{i+1}$, then let $F'_i \in \mathcal{F}(C)$ be such that vectors $\{u_1,...,u_{i-1}, u_{i+1} \}$ are a basis of $F'_i$ and $F_{i-1} \subset F'_i$. We have: $F'_i \neq F_i$ and $F'_i \subset F_{i+1}$, thus the set $S' = S_F \cup  \{\bar{B}_M(F'_i) \}$ is such that $[S']$ is a $d-1$-dimensional simplex of $[B]^\star$ and $[S_F] = [S] \cap [S']$;
	\item $p = \bar{B}_M(F_1) = b \in B$. There exists $b' \in B$ such that $b' \in F_2$ and $b' \neq b$ (using the proof of proposition \ref{prop:simplex}). Let $F'_1$ be the edge of $C$ such that $b' \in F'_1$. We have $\{F'_1, F_2, ..., F_{d-1} \} \in \mathcal{F}^\star(B,C)$. Hence $S' = \{ \bar{B}_M(F'_1) \} \cup S_F$ is such that $[S']$ is a $d-1$-dimensional simplex of $[B]^\star$ and $[S_F] = [S] \cap [S']$.
\end{itemize}
Finally, each simplex of $[B]^\star$ has one of its facets which is included in $\partial C$ the boundary of $C$ and shares all its other facets with other simplices of $[B]^\star$. Therefore the boundary of $[B]^\star$ is included in $\partial C$.
$\square$  \vspace{0.5 cm}

In 2 dimensions (see examples on Figure \ref{fig:cresistar2D}), the simplices of the c-resistar are segments $[b, \bar{B}]$, therefore, the number of simplices equals the number of boundary points. 

In 3 dimensions (see an example on Figure \ref{fig:cresistar3D}), the simplices are triangles $[b, \bar{B}_M(F_2), \bar{B}]$, where $F_2$ is a 2D face of the cube including boundary point $b$. For each boundary point, there are two such faces, therefore, the number of simplices is twice the number of boundary points.

More generally, for $2 \leq i \leq d-1$, there are $d - i +1$ faces of dimensionality $i$ including a face of dimensionality $i-1$, therefore, the number of simplices in a $d$-dimensional cube is $(d-1)!$ times its number of boundary points. The number of simplices increases thus very rapidly with the dimensionality. However, this is not a problem in our perspective because, as shown further, it is possible to compute efficiently the classification of a point by a c-resistar without considering the simplices explicitly.

\subsubsection{Classification function.}
%-------------------------------------------

Propositions \ref{prop:simplex} and \ref{prop:border} imply that the c-resistar is a $d-1$-dimensional set dividing the cube $C$ into several connected sets that we call classification sets. Definition \ref{def:Cj} expresses these sets as the union of polytopes sharing vertices with faces of $C$ and with the simplices of $[B]^\star$. The following propositions establish their properties. Finally these sets are used to define the resistar classification function.

%Proposition \ref{prop:respectc} states important properties of sets $C_j$ that will be used to define the classification function.

\begin{definition}
Let $C$ be a $d$-dimensional cube of the grid, such that $B = B_M(C) \neq \emptyset$. Let $\mathcal{F}(C)$ be the set of all the faces of cube $C$ (including $C$ itself) and $\mathcal{H}(B, C)$ be the set of the faces of $C$ without boundary points:
\begin{align}
	\mathcal{H}(B,C) = \left\{ F \in \mathcal{F}(C), B_M(F) = \emptyset \right\}. %\cup \left\{ [v_-(b), v_+(b)], b \in B \right\},
\end{align}

Let $\mathcal{D}(B,C)$ be the set of connected components of $\mathcal{H}(B,C)$.

For all $D \in \mathcal{D}(B,C)$, the classification set $[B]^{\mathcal{D}  \star}(D)$ associated to $D$ by resistar $[B]^\star$ in $C$, is defined as follows:
	\begin{align}
	[B]^{\mathcal{D}  \star}(D) =  \bigcup_{ \{ F_h,..,F_{d-1} \} \in \mathcal{F}^{\mathcal{D}  \star}(D,C) }\left[\mathcal{V}(F_h) \cup \{\bar{B}_M(F_{h+1}),...,\bar{B}_M(F_{d-1}) \} \cup \{\bar{B} \} \right],
	\end{align}
	with:
	\begin{align}
	\{ F_h,..,F_{d-1} \} \in \mathcal{F}^{\mathcal{D}  \star}(D,C) \Leftrightarrow
	\begin{cases}
	F_h \in \mathcal{F}_h(C), F_h \subset D, 0 \leq h \leq d-1, \\
	F_i \in \mathcal{F}_i(C), F_{i-1} \subset F_{i}, h+1 \leq i \leq d-1,\\
	B_M(F_{h+1}) \neq \emptyset,
	\end{cases}
	\label{eq:Ck}
\end{align}
where $i$ and $h$ are integers, $\mathcal{V}(F_h)$ is the set of vertices of face $F_h$. 
\label{def:Cj}
\end{definition}

\begin{figure}
 	\centering
 	\begin{tabular}{cc}
 		\includegraphics[width=6 cm]{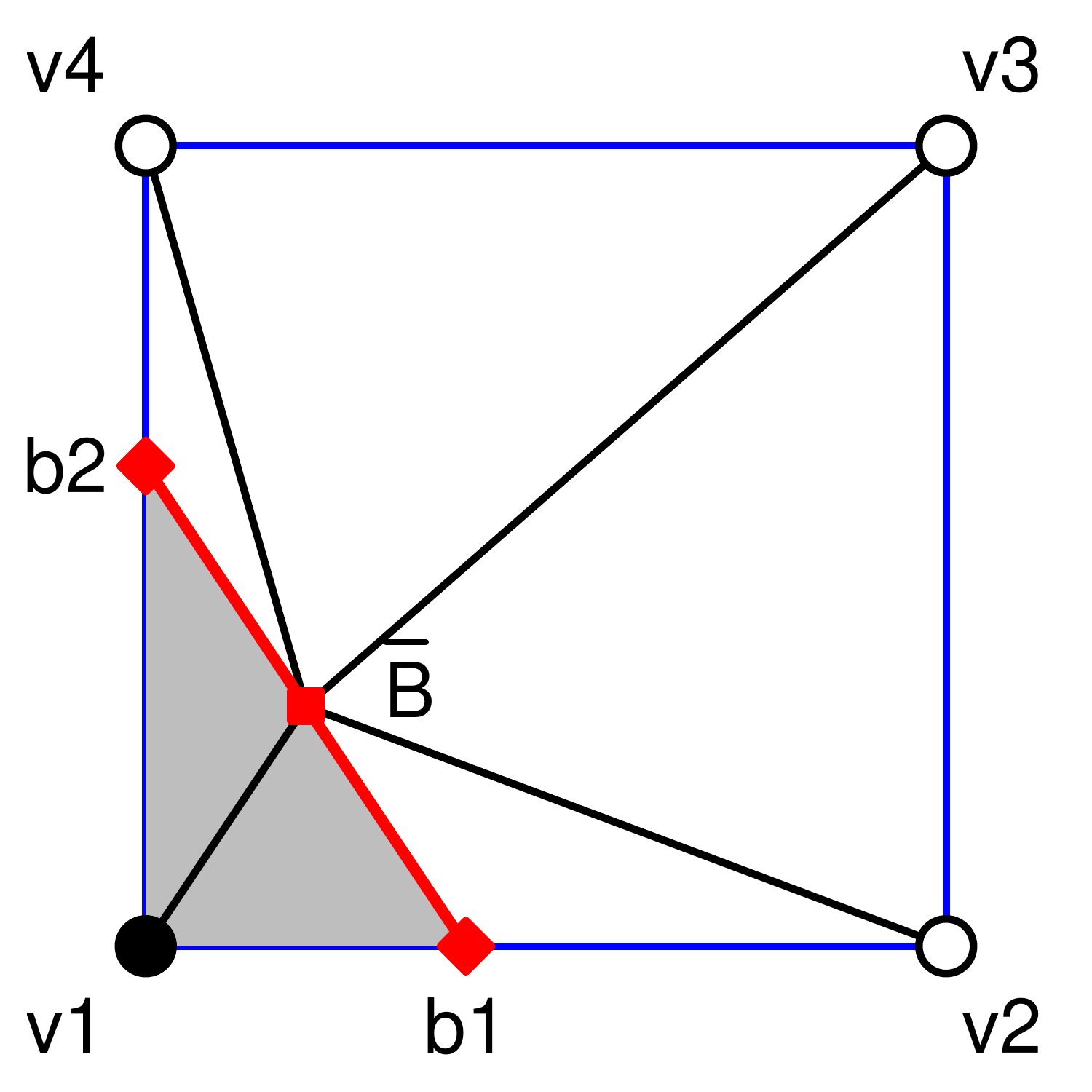} & 
 		\includegraphics[width= 6 cm]{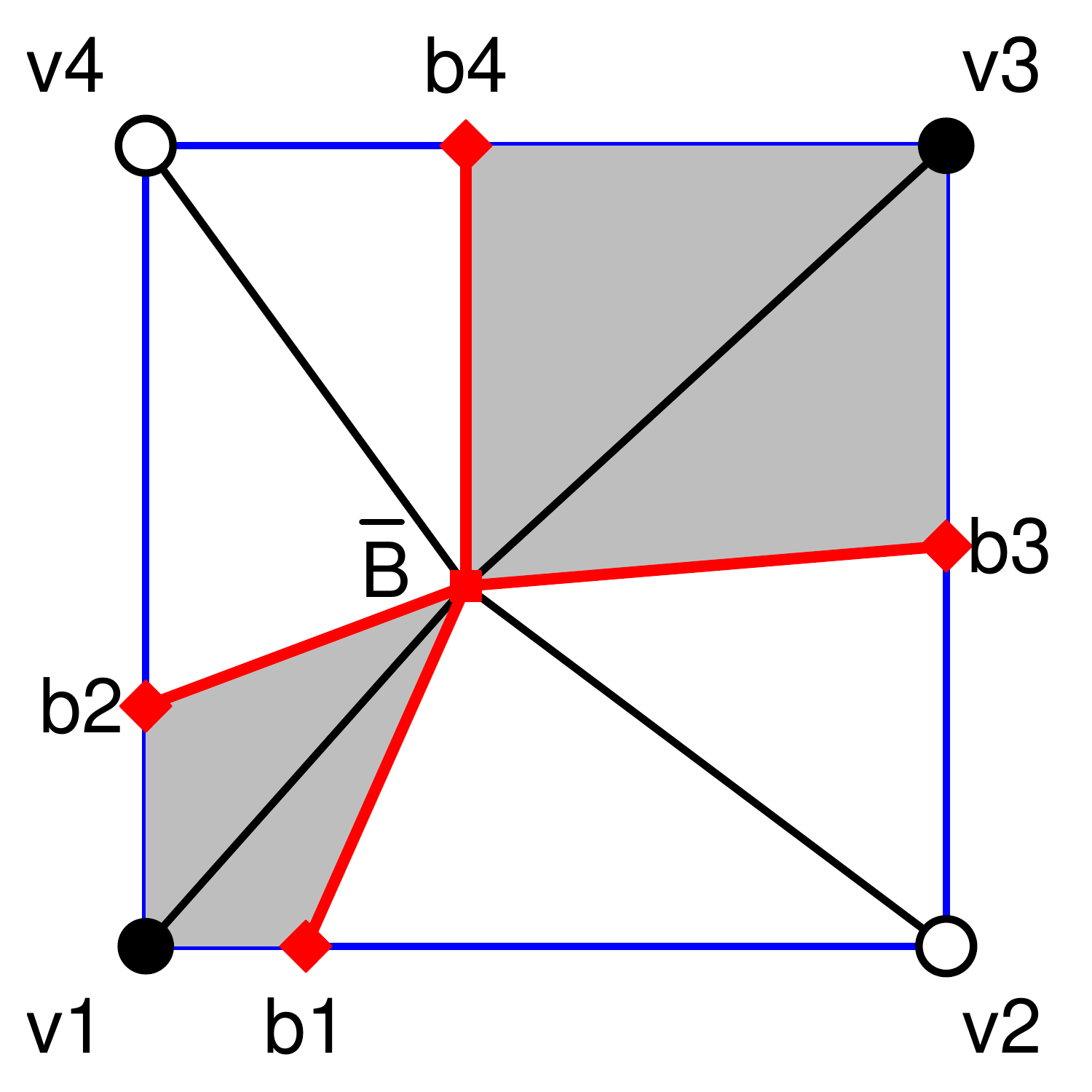} \\
		(a) & (b)
 	\end{tabular}
 	\caption{Examples of sets of connected faces without boundary points $D$ and classification sets $[B]^{\mathcal{D}\star}(D)$ from c-resistar in 2 D cubes. Panel (a): $\mathcal{D}(B,C)$ includes two sets: $D_1 = \{ v_1 \}$, $D_2 =  [v_2, v_3] $$ \cup [v_3, v_4] $. The classification sets are:  $[B]^{\mathcal{D}  \star}(D_1) =  [ v_1, b_1, \bar{B }] $ $\cup [v_1, b_2, \bar{B }] $ and $[B]^{\mathcal{D}  \star}(D_2) = [v_2, b_1, \bar{B }]$$ \cup [v_2, v_3, \bar{B }] $$ \cup [v_3, v_4, \bar{B }] $$\cup [v_4, b_2, \bar{B }] $. Panel (b):  $\mathcal{D}(B,C)$ includes four sets: $D_1 = \{ v_1 \}$ , $D_2 = \{ v_2 \}$ , $D_3 = \{ v_3 \}$ , $D_4 = \{  v_4 \}$. The classification sets are: $[B]^{\mathcal{D}  \star}(D_1) =  [v_1, b_1, \bar{B }] $$\cup [v_1, b_2, \bar{B }] $, $[B]^{\mathcal{D}  \star}(D_2) =  [v_2, b_1, \bar{B }] $$\cup [v_2, b_3, \bar{B }] $, $[B]^{\mathcal{D}  \star}(D_3) =  [v_3, b_3, \bar{B }] $$\cup [v_3, b_4, \bar{B }] $, $[B]^{\mathcal{D}  \star}(D_4) =  [v_4, b_4, \bar{B }] $$\cup [v_4, b_2, \bar{B }] $. }
 	\label{fig:cresistar2DC}
 \end{figure} 

Figure \ref{fig:cresistar2DC} shows examples of c-resistar classification sets in 2D.

\begin{proposition}
For all $D \in \mathcal{D}(B,C)$, $D \subset [B]^{\mathcal{D}  \star}(D)$, $[B]^{\mathcal{D}  \star}(D)$ is a $d$-dimensional connected set and $[B]^{\mathcal{D}\star}(D) \cap ([B]^\star - \partial [B]^\star)$ is the boundary of $[B]^{\mathcal{D} \star}(D)$ in $C - \partial C$.
\label{prop:respectc}
\end{proposition}

\textbf{Proof.} Obviously, $D \subset [B]^{\mathcal{D}  \star}(D)$, by definition. Indeed, for any face $F_h \in \mathcal{F}_h(C)$, such that $F_h \subset D$, we have $[\mathcal{V}(F_h)] = F_h \subset [B]^{\mathcal{D}  \star}(D)$ directly from the definition of $[B]^{\mathcal{D} \star}(D)$.

Let $P = \mathcal{V}(F_h) \cup \{\bar{B}_M(F_{h+1}),...,\bar{B}_M(F_{d-1}) \} \cup \{\bar{B} \}$, with $\{ F_h, ...,F_{d-1} \} \in \mathcal{F}^{\mathcal{D} \star}(D, C)$. $[P]$ is a polytope of dimensionality $d$ because $F_h$ is of dimensionality $h$ and the set $P - \mathcal{V}(F_h)$ includes $d-h$ affinely independent points (see proposition \ref{prop:simplex}), which are not located in $F_h$. 

Let $P_F \subset P$ such that $[P_F]$ is a facet of $[P]$. There are two possibilities:
\begin{itemize}
\item $P_F = P - \{ p\}$, with $p \in \{\bar{B}_M(F_{h+1}),...,\bar{B}_M(F_{d-1}) \} \cup \{\bar{B} \}$. The following cases occur:
\begin{itemize}
	\item $p = \bar{B}$. Then $[P_F]$ is included in $F_{d-1}$, therefore $\left[P_F\right] \subset \partial C$;
	 \item $p =\bar{B}_M(F_i)$, with $h+1 \leq i \leq d-1$, then two cases occur again: 	
	\begin{itemize}
		\item $h+1 < i$. As shown in the proof of proposition \ref{prop:border}, there exists $F'_i \in \mathcal{F}_i(C)$ such that $F'_i \neq F_i$, $F_{i-1} \subset F'_i$, and $F'_i \subset F_{i+1}$. Moreover, because $i > h+1$, $F_{h+1} \subset F_{i-1} \subset F'_i$. Hence, $[P']$, with $P' = P_F \cup \{ \bar{B}_M(F'_i) \}$, is a polytope of $[B]^{\mathcal{D} \star}(D)$ and $[P_F] = [P] \cap [P']$.
		\item $i = h+1$. Let $F'_{h+1} \in \mathcal{F}_{h+1}(C)$ be such that $F_h \subset F'_{h+1}$ and $F'_{h+1} \subset F_{h+2}$. There are two possibilities:
		
		\begin{itemize}
			\item $B_M(F'_{h+1}) = \emptyset$, then $P' = \mathcal{V}(F'_{h+1}) \cup \{ \bar{B}_M(F_{h+2}),...,\bar{B}_M(F_{d-1}) \} \cup \{\bar{B} \}$ is such that $[P']$ is a polytope of $[B]^{\mathcal{D} \star}(D)$ because $F'_{h+1}$ is connected to $F_h$ and $[P_F] = \left[P - \bar{B}_M(F_{h+1})\right] = [P] \cap [P']$.
			\item $B_M(F'_{h+1}) \neq \emptyset$, then $P' = P_F  \cup \{\bar{B}_M(F'_{h+1}) \}$ is such that $[P']$ is a polytope of $[B]^{\mathcal{D} \star}(D)$ and $[P_F] = [P] \cap [P']$.
		\end{itemize}
	\end{itemize}
	\end{itemize}
	
		\item The $d-1$-dimensional face is obtained by removing vertices in $\mathcal{V}(F_h)$. Thus:
		
		\begin{itemize}
			\item If $h = 0$, $\mathcal{V}(F_h)$ is a single vertex $v\in D$, and the edge $F_1 = [v, v']$ is such that $B_M(F_1) = \{ b \}$ . Then $P_F = P - \{v\} = \{ \bar{B}_M(F_1), ..., \bar{B}_M(F_{d-1}) \} \cup \{  \bar{B} \}$ and $[P_F] \subset [B]^\star$. Moreover,  $P' = \{ v'\} \cup P_F$ is such that $[P_F] = [P] \cap [P']$ and there exists $D' \in \mathcal{D}(B,C), D' \neq D$ such that $v' \in D'$ and $[P'] \subset [B]^{\mathcal{D} \star}(D')$.
			\item If $h > 0$, 	$ P_F = \mathcal{V}(F_{h-1}) \cup \{ \bar{B}_M(F_{h+1}),..., \bar{B}_M(F_{d-1}) \} \cup \{ \bar{B} \}$, with $F_{h-1} \in \mathcal{F}_{h-1}(F_h)$. Let $F'_{h} \in \mathcal{F}_h(C)$ be such that $F_{h-1} \subset F'_h$, $F'_h \subset F_{h+1}$ and $F'_h \neq F_h$. There are two possibilities:
\begin{itemize}
	\item $B_M(F'_h) \neq \emptyset$, then $P' = P_F \cup \{ \bar{B}_M(F'_{h}) \}$ is such that $[P']$ is a polytope of $[B]^{\mathcal{D} \star}(D)$ and  $[P_F] = [P] \cap [P']$;
	\item $B_M(F'_h) = \emptyset$, then $F'_h \subset D$, because $F_h$ and $F'_h$ are connected by sharing face $F_{h-1}$ and $P' = \mathcal{V}(F'_{h}) \cup \{ \bar{B}_M(F_{h+1}), ..., \bar{B}_M(F_{d-1}) \} \cup \{\bar{B} \}$ is such that $[P']$ is a polytope of $[B]^{\mathcal{D} \star}(D)$ and  $[P_F] = [P] \cap [P']$ (because $F_{h-1} \subset F'_{h}$).
\end{itemize}
		\end{itemize}

\end{itemize}
Finally, all the polytopes $[\mathcal{V}(F_h) \cup \{\bar{B}_M(F_{h+1}),...,\bar{B}_M(F_{d-1}) \} \cup \{\bar{B} \}]$ of $[B]^{\mathcal{D} \star}(D)$ such that $h > 0$ have all their facets shared with another polytope of $[B]^{\mathcal{D} \star}(D)$ except one facet which is included in $\partial C $; the polytopes such that $h = 0$ have all their facets shared with another polytope of $[B]^{\mathcal{D} \star}(D)$ except one which is included in $\partial C $ and one which is included in $[B]^\star$.
Therefore, $[B]^{\mathcal{D} \star}(D)$ is a connected set and its boundary in $C - \partial C$ is included in $[B]^\star$.

Moreover, the set of facets included in $[B]^\star$ from polytopes such that $h = 0$ is:
\begin{align}
E =  \bigcup_{\{F_1,...,F_{d-1} \} \in \mathcal{F}^\star(B,C), F_1 \cap D \neq \emptyset}[\{\bar{B}_M(F_1),...,\bar{B}_M(F_{d-1}\} \cup \{ \bar{B} \} ].
\end{align}
We have $[B]^\star \cap [B]^{\mathcal{D} \star}(D) \subset E$. Indeed, for any set $\{F_h,...,F_{d-1} \} \in \mathcal{F}^{\mathcal{D}\star}(D,C)$, such that $h > 0$, there exists $\{F_1,...,F_{d-1} \} \in \mathcal{F}^\star(B,C)$, such that $F_1 \subset F_{h+1} $, $F_1 \cap D \neq \emptyset$ and $\{F_{h+1},...,F_{d-1} \} \subset \{F_1,...,F_{d-1} \}$. Therefore, for all polytopes $[P]$ such that $h > 0$ the part of the boundary of $[P]$ which is included in $[B]^\star$ is also included in $E$. Moreover, $E \subset [B]^\star \cap [B]^{\mathcal{D} \star}(D)$ by definition. Therefore, $E = [B]^\star \cap [B]^{\mathcal{D} \star}(D)$.

Moreover, $E - (E \cap \partial C)$ is the boundary of  $[B]^{\mathcal{D} \star}(D)$ in $C - \partial C$, hence the boundary of $[B]^{\mathcal{D} \star}(D)$ in $C - \partial C$ is $[B]^{\mathcal{D}\star}(D) \cap ([B]^\star - \partial[B]^\star)$.
$\square$  %\vspace{0.5 cm}

\begin{proposition}
The union of the classification sets defined by the c-resistar $[B]^\star$ in cube $C$ is cube $C$ itself:
	\begin{align}
	C = \bigcup_{D \in \mathcal{D}(B,C)} [B]^{\mathcal{D} \star}(D).
	\label{eq:union}
\end{align}
\label{prop:union}
\end{proposition}

\textbf{Proof.}
The proof of proposition \ref{prop:respectc} shows that for all polytopes $[P] \in [B]^{\mathcal{D} \star}(D)$, a facet of $[P]$ which is not in $\partial C$ is either shared with another polytope of $[B]^{\mathcal{D} \star}(D)$ or with another polytope of $[B]^{\mathcal{D} \star}(D')$, with $D' \neq D$.
Therefore, because the sets $[B]^{\mathcal{D} \star}(D)$ are of dimensionality $d$, this union is $C$ itself.
$\square$  %\vspace{0.5 cm}

\begin{proposition}
For $(D,D') \in \mathcal{D}(B,C)^2$, $D' \neq D$, $([B]^{\mathcal{D} \star}(D) \cap [B]^{\mathcal{D} \star}(D')) \subset [B]^\star$.
\label{prop:inter}
\end{proposition}

\textbf{Proof}
Consider $(D,D') \in \mathcal{D}(B,C)^2$, $D \neq D'$. Consider polytope $[P] \in [B]^{\mathcal{D} \star}(D)$ with $P = \mathcal{V}(F_h) \cup \{\bar{B}_M(F_{h+1}),..., \bar{B}_M(F_{d-1}) \} \cup \{ \bar{B} \}$ and polytope $[P'] \in [B]^{\mathcal{D} \star}(D')$ with $P' = \mathcal{V}(F'_h) \cup \{\bar{B}_M(F'_{h+1}),..., \bar{B}_M(F'_{d-1}) \} \cup \{ \bar{B} \}$. We have: $[P] \cap [P'] = [P \cap P']$ and $F_h \cap F'_{h'} = \emptyset$. Therefore the intersection between $[P]$ and $[P']$ is a simplex of dimensionality at most $d-1$, of vertices the points $\bar{B}_M(F_i)$ such that there exists $(b, b') \in B_M(F_i)^2$ with ($v_+(b) \in D$ or $v_-(b) \in D$) and ($v_+(b') \in D'$ or $v_-(b') \in D'$). Therefore $([B]^{\mathcal{D} \star}(D) \cap [B]^{\mathcal{D} \star}(D')) \subset [B]^\star$. 
$\square$ \vspace{0.5 cm}

These properties of the classification sets $[B]^{\mathcal{D} \star}(D)$ guarantee that for any point $x \in C - [B]^\star$, there exists a unique set $D$ such that $x \in [B]^{\mathcal{D}\star}(D)$. This leads to the definition of the resistar classification.
 
\begin{definition}
Let $C$ be a cube of the grid such that $B = B_M(C)$, The resistar classification function $[B]^\star(.)$, from $C$ to $\{-1, 0, +1\}$, is defined for $x \in C$ as follows:
\begin{itemize}
	\item If $B = \emptyset$, then $[B]^\star(x) = \mathcal{M}(v)$, $v \in \mathcal{V}(C)$;
	\item Otherwise, let $\mathcal{D}(B,C)$ be the set of connected faces of $C$ without boundary points and, for $D \in \mathcal{D}(B,C)$, let $[B]^{\mathcal{D} \star}(D)$ be the classification associated to $D$ by the c-resistar $[B]^\star$.
	\begin{itemize}
	\item If $x \in [B]^\star$ then $[B]^\star(x) = 0$, 
	\item Otherwise there exists a single set $D \in \mathcal{D}(B,C)$ such that $x \in [B]^{\mathcal{D} \star}(D)$ and $[B]^\star(x) = \mathcal{M}(v)$, $v \in \mathcal{V}(D)$. 
	\end{itemize}
\end{itemize}
\label{def:classifc}
\end{definition}

This classification function is consistent with the classification of the vertices of $C$ by $\mathcal{M}$ because, for any vertex $v \in \mathcal{V}(D)$, $D \in \mathcal{D}(B,C)$, $[B]^\star(v) = \mathcal{M}(v)$, and any point $x \in [B]^{\mathcal{D} \star}(D)$ is on the same side of $[B]^\star$ as $v$, since $[B]^\star \cap [B]^{\mathcal{D} \star}(D)$ includes the boundary of $[B]^{\mathcal{D} \star}(D)$ in $C - \partial C$.

\subsubsection{Classification algorithm.}
%-------------------------------------------

\begin{figure}
 	\centering
 	\begin{tabular}{cc}
 		\includegraphics[width=6 cm]{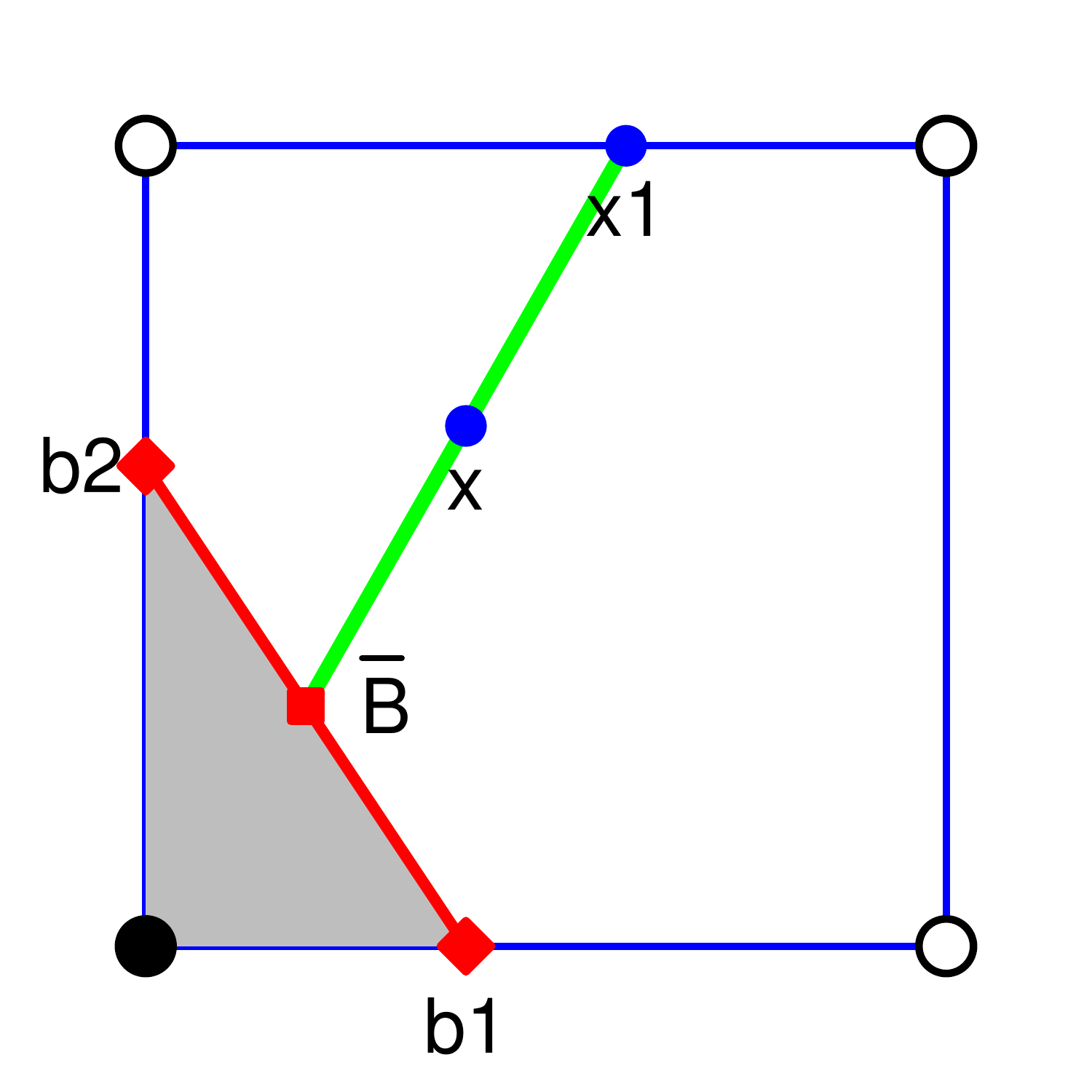} & 
 		\includegraphics[width= 6 cm]{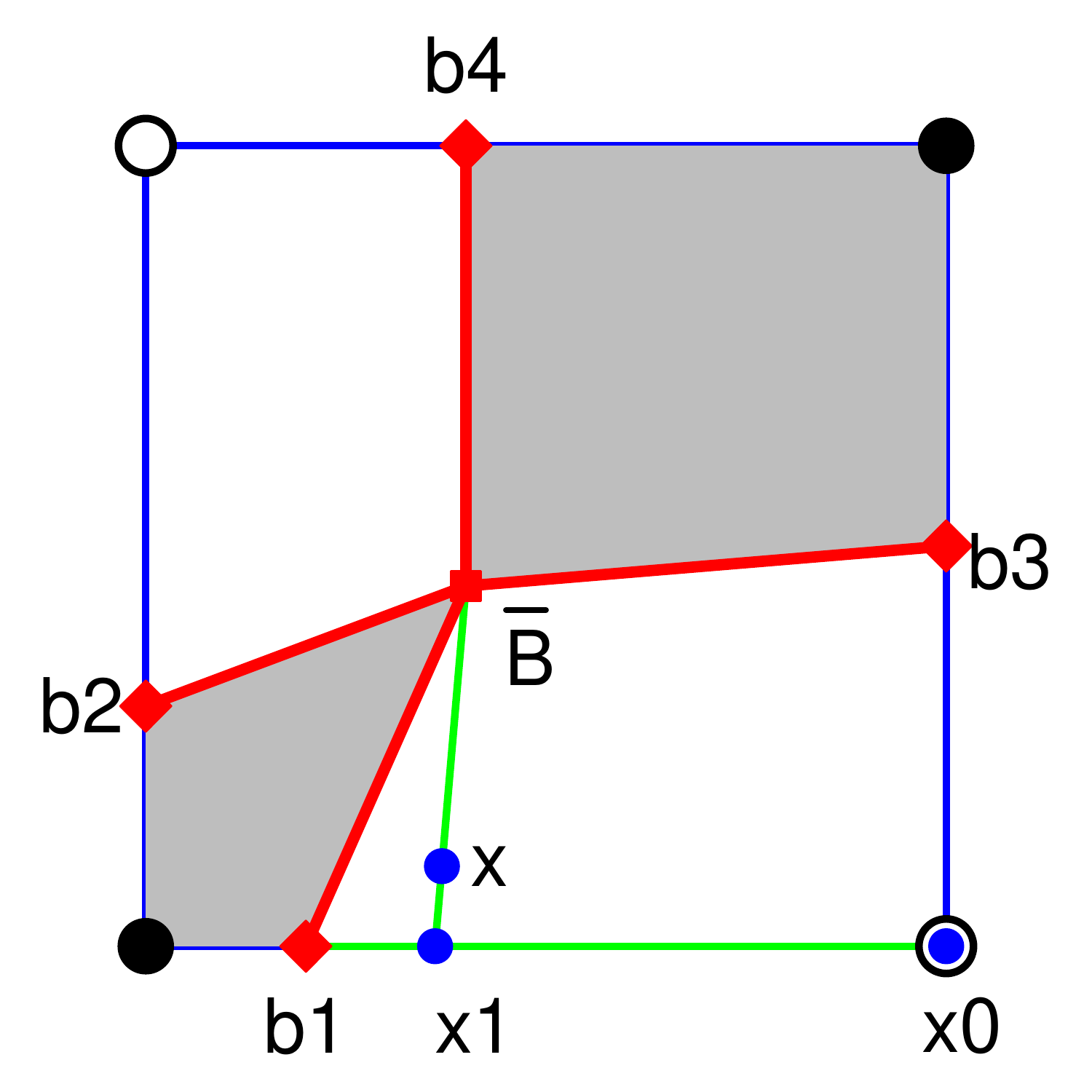} \\
		(a) & (b)
 	\end{tabular}
 	\caption{Illustration of algorithm \ref{Algo:cFaceProject} for resistar classification in 2D. Panel (a): the point $x_1 = \vec{r}(\bar{B }, x) \cap \partial C$ is in a face $F$ such that $F \cap B = \emptyset$, and $\mathcal{M}(v) = +1$, for $v \in \mathcal{V}(F)$ therefore $[B]^\star(x) = +1$ (the white vertices $v$  of $C$ are such that $\mathcal{M}(v) = +1$). Panel (b): the point $x_1 = \vec{r}(\bar{B }, x) \cap \partial C$ is in a face $F$ such that $F \cap B \neq \emptyset$, $x_0 = \vec{r}(b_1, y_1) \cap \partial F$ is a vertex of $C$ (a face of dimensionality 0), such that $\mathcal{M}(x_0) = +1$ therefore $[B]^\star(x) = +1$.}
 	\label{fig:cresistar2DClassif}
 \end{figure}

Algorithm \ref{Algo:cFaceProject} takes as input a point $x$ of cube $C$ and, if $B = B_M(C)$ is empty, it returns the classification of a vertex of $C$ by $\mathcal{M}$. Otherwise, if $x$ is not equal to $\bar{B}$, it computes $x_{d-1} = \vec{r}(\bar{B},x) \cap \partial C$, where $\vec{r}(\bar{B},x)$ is the ray from $\bar{B}$ in the direction of $x$ and $\partial C$ is the boundary of $C$. $x_{d-1}$ is located in a facet $F_{d-1}$ of $C$ and algorithm \ref{Algo:cFaceProject} repeats the same operations for $x_{d-1} \in F_{d-1}$. When it reaches a face $F_i$ without boundary points, the algorithm returns the classification of a vertex of $F_i$, or when  $x_i = \bar{B}_M(F_i)$ it returns $0$ (see examples on Figure \ref{fig:cresistar2DClassif}). 

\begin{algorithm}[H]
%\dontprintsemicolon
\KwIn{$B$ set of boundary points in $d$ dimensional cube $C$, $x \in C$ point to classify.}
	$F_d \leftarrow C$;
	$x_d \leftarrow x$;
	$B_d \leftarrow B$;
	$i \leftarrow d$;\\
	\While{$i \geq 0$}
	{\If{$B_i = \emptyset$}{\Return $\mathcal{M}(v), v \in \mathcal{V}(F_i)$;}
	\If{$x_i = \bar{B}_i  $}{\Return $0$}
	$x_{i-1} \leftarrow \vec{r}(\bar{B}_i  , x_i) \cap \partial F_i$;\\
	$F_{i-1} \leftarrow F \in \mathcal{F}_{i-1}(F_i),  x_{i-1} \in F$;\\
	$B_{i-1} \leftarrow F_{i-1} \cap B_i$;\\
	$i \leftarrow i-1$}
\label{Algo:cFaceProject}
\caption{Classification of point $x$ in cube $C$ by c-resistar $[B]^\star$.}
\end{algorithm}

Algorithm \ref{Algo:cFaceProject} always terminates, because the dimensionality of face $F_i$ decreases of 1 at each step, and in the worst case the algorithm reaches a face of dimensionality 0, which, by hypothesis, cannot include any boundary point. %Proposition \ref{prop:classifc} states that algorithm \ref{Algo:cFaceProject} returns the classification $[B]^\star(x)$ of point $x \in C$.

\begin{proposition}
Algorithm \ref{Algo:cFaceProject} applied to $x \in C$ returns $[B]^\star(x)$, as specified in definition \ref{def:classifc}.
\label{prop:classifc}
\end{proposition}

\textbf{Proof.} Consider $x \in C$.
\begin{itemize} 
	\item If $x \in [B]^\star$, let $S = \{ \bar{B}_M(F_1),...,\bar{B}_M(F_{d-1}) \} \cup \{ \bar{B} \}$, with $\{F_1,...,F_{d-1} \} \in \mathcal{F}^\star(B,C)$, such that $x \in [S]$. At each step of algorithm \ref{Algo:cFaceProject}, either $x_i = \bar{B}_M(F_i) $ and then the algorithm returns $0$, or $x_{i-1} \in [\bar{B}_M(F_1),...,\bar{B}_M(F_{i-1}) ]$. Therefore, in the worst case, the algorithm reaches $x_1 \in [\bar{B}_M(F_1)] = b $, $b \in B$, in which case the only possibility is $x_1 = b$, and the algorithm returns $0$. Therefore if $x \in [B]^\star$, algorithm \ref{Algo:cFaceProject} returns 0.
	\item If $x \in C - [B]^\star$, let $\{x_d = x, .., x_h \}$ be the values of $x_i$ at the successive steps of algorithm \ref{Algo:cFaceProject}, $h$  being the dimensionality of the face $F_h$ such that $F_h \cap B = \emptyset$ at the last step of the algorithm. There exists a unique set of connected faces without boundary points $D$ such that $F_h \subset D$ thus $x_h \in D$ and $x_h \in [B]^{\mathcal{D} \star}(D)$. For $h+1 \leq i \leq d$ the rays $\vec{r}(\bar{B}_M(F_i)  , x_i)$  do not cross any simplex of $[B]^\star$, thus the segments $[x_i, x_{i-1}]$ for $h+1 \leq i \leq d$ do not cross $[B]^\star$ either. Therefore, because the boundary of $[B]^{\mathcal{D} \star}(D)$ in $C- \partial C$ is included in $[B]^\star$, $x = x_d \in [B]^{\mathcal{D} \star}(D)$. Therefore, algorithm \ref{Algo:cFaceProject} returns $[B]^\star(x)$.	
\end{itemize}
$\square$ %\vspace{0.5 cm}

Actually, it can easily be shown that $P = \mathcal{V}(F_h) \cup \{\bar{B}_M(F_{h+1}),..., \bar{B}_M(F_{d-1})\} \cup \{ \bar{B}\}$, with the faces $F_i$ defined by algorithm \ref{Algo:cFaceProject} is such that $[P]$ a polytope of $[B]^{\mathcal{D} \star}(D)$ and $x \in [P]$.

It appears finally that, even if the set $[B]^\star$ includes a large number of simplices, the classification algorithm requires at most $d$ relatively light computations (selecting boundary points in a face, computing their barycentre, projecting a point on the boundary of the face). Section \ref{sec:memory} presents a modification of this algorithm with a better management of the memory space.

%----------------------------------------------------
\subsection{c-resistar approximation on the grid.} 
%----------------------------------------------------
The regular grid $G$ comprises $n_G^d$ points covering $X$ and its facets. The values of the coordinates of the grid points are taken in $\left\{ 0, \frac{1}{n_G-1}, \frac{2}{n_G-1}..., 1  \right\}$. $\mathcal{F}_i(G)$ denotes all the $i$-dimensional cubes or cube faces of $G$ and $\mathcal{F}(G)$ denotes the set of all cubes or cube faces of $G$.

\subsubsection{Definition.}
%----------------------------------------

The definition of the c-resistar approximation of $M$ on grid $G$ comes directly from the definition of the c-resistars in the cubes. 

\begin{definition}
The c-resistar approximation of $M$ on grid $G$, denoted $[B_M(G)]^\star$, $B_M(G)$ being the set the boundary points from of all the edges of the cubes of the grid, is the union of the c-resistars approximating $M$ in the cubes of $G$:

\begin{align}
	 [B_M(G)]^{\star} = \bigcup_{C \in \mathcal{F}_d(G), B_M(C) \neq \emptyset} [B_M(C)]^\star.
\end{align}
%where $[B_M(C)]^\star = \emptyset$ if $B_M(C) = \emptyset$.
\end{definition}

\begin{proposition}
The c-resistar approximation of $M$ on grid $G$ is a set of $d-1$-dimensional simplices and its boundary is included in the boundary of $X$.
\label{prop:bordercg}
\end{proposition}

\textbf{Proof.}
$[B_M(G)]^{\star}$ is a set of $d-1$-dimensional simplices as the union of sets of $d-1$-dimensional simplices (see proposition \ref{prop:simplex}). 

Let $C$ be a cube of $G$, with $B = B_M(C) \neq \emptyset$, let $S = \{\bar{B}_M(F_1),...,\bar{B}_M(F_{d-1}) \} \cup \{ \bar{B} \}$, such that $[S]$ is a $d-1$-dimensional simplex in $[B]^\star$. The facet $[S_F] = [S - \{ \bar{B}_M(C) \}]$ of $S$ is located in facet $F_{d-1}$ of $C$. Two cases occur:
\begin{itemize}
	\item There exists a cube $C'$, with $B_M(C') = B'$, sharing facet $F_{d-1}$ with $C$. Then the simplex $[S'] = [S_F \cup \{ \bar{B'} \}]$ is such that $[S'] \in [B_M(C')]^\star$ and $[S_F] = [S] \cap [S']$;
	\item Otherwise, $[S_F]$ is included in the boundary of $X$.
\end{itemize}
Therefore,  taking proposition \ref{prop:simplex} into account, all the simplices of $[B_M(G)]^{\star}$ share all their facets with other simplices of $[B_M(G)]^{\star}$, except the facets which are in the boundary of $X$.
$\square$ % \vspace{0.5 cm}

\subsubsection{Classification by the c-resistar approximation on the grid.}
%----------------------------------------
The classification sets defined by the c-resistar approximation on the grid are derived from the classification sets in the grid cubes.

\begin{definition}
Let $B_G = B_M(G)$, $\mathcal{H}(B_G,G) = \left\{ F \in \mathcal{F}(G), B_M(F) = \emptyset \right\}$ and $\mathcal{D}(B_G,G) $ be the connected components of $\mathcal{H}(B_G,G)$.
 
For $D \in \mathcal{D}(B_G,G)$ the classification set $[B_G]^{\mathcal{D} \star}(D)$ defined by $[B_G]^\star$, the c-resistar approximation of $M$ on grid $G$, is the set:
\begin{align}
	[B_G]^{\mathcal{D} \star}(D) = \bigcup_{C \in \mathcal{F}_d(G)} [B_M(C)]^{\mathcal{D} \star}(D \cap C), 
\end{align}
where:
\begin{align}
	[B_M(C)]^{\mathcal{D} \star}(D \cap C) = 
	\begin{cases}
	\emptyset, \mbox{ if } D \cap C = \emptyset,\\
	C, \mbox{ if } D \cap C = C,
	\end{cases}
\end{align}
and follows definition \ref{def:Cj} otherwise.

\end{definition}

\begin{proposition}
 For all $D \in \mathcal{D}(B_G,G) $, $D \subset [B_G]^{\mathcal{D} \star}(D)$, $[B_G]^{\mathcal{D} \star}(D)$ is connected and its boundary in $X - \partial X$ is $[B_G]^{\mathcal{D} \star}(D) \cap ([B_G]^{\star} - \partial [B_G]^{\star}$).
\label{prop:cseparationG}
\end{proposition}

\textbf{Proof.}
\begin{itemize}
	\item $D \subset [B_G]^{\mathcal{D} \star}(D)$ by definition.

 \item All the polytopes of $[B_G]^{\mathcal{D} \star}(D)$ defined from face $F_h \in D$ are connected to each other because they share face $F_h$. Because $D$ is a connected set by definition, the polytopes defined from all faces $F_h \in D$ are connected to each other through the connections between faces $F_h$. Therefore $[B_G]^{\mathcal{D} \star}(D)$ is a connected set. 

\item All the polytopes of $[B_G]^{\mathcal{D} \star}(D)$ are of dimensionality $d$, because they are either cubes without boundary points, or polytopes of a classification set in a cube, which are all of dimensionality $d$.

\item Consider $D \in \mathcal{D}(B_G,G)$ and a cube $C$ such that $D \cap C \neq \emptyset$. 
\begin{itemize}
\item If $C \subset D$, $[B_M(C)]^{\mathcal{D} \star}(C \cap D) = C$ and for any $d-1$-dimensional face $F_{d-1}$ of $C$ which is shared with another cube $C' \in \mathcal{F}_d(G)$, there exists a polytope $[P'] \subset [B_M(C')]^{\mathcal{D} \star}(D \cap C')$ such that $F_{d-1} \subset [P']$. Indeed, if $C' \subset D$, then $ [B_M(C')]^{\mathcal{D} \star}(D \cap C') = C'$ and $F_{d-1} \subset C \cap C'$ by hypothesis. Otherwise, we have $\{ F_{d-1} \} \in \mathcal{F}^{\mathcal{D} \star}(D \cap C', C')$ because $B_M(F_{d-1}) = \emptyset$;
\item If $D \cap C \neq C$, let $P = \mathcal{V}(F_h) \cup \{ \bar{B}_M(F_{h+1}),..., \bar{B}_M(F_{d-1})  \} \cup \{ \bar{B}_M(C) \}$,  with $\{ F_h,..., F_{d-1} \} \in \mathcal{F}^{\mathcal{D} \star}(D \cap C, C) $. $[P]$ is a polytope included in $[B_G]^{\mathcal{D} \star}(D)$. Let $P_F = P - \{ \bar{B}_M(C) \}$. $[P_F]$ is a face of $[P]$ which is included in facet $F_{d-1}$ of $C$. If there exists a cube $C' \in \mathcal{F}_d(G)$, $C' \neq C$, such that $F_{d-1} \subset C'$, then:
\begin{itemize}
\item if $C' \subset D$, $[P_F] = F_{d-1}$ and $F_{d-1}$ is shared by $[P]$ and $C'$;
\item if $C' \cap D \neq C'$, then $B_M(C') \neq \emptyset$. The set $P' = P_F \cup \{ \bar{B}_M(C') \}$ is such that $[P'] \subset [B_M(C')]^{\mathcal{D} \star}(D \cap C')$ and $[P_F] = [P] \cap [P']$. 
\end{itemize}
As shown in the proof of proposition \ref{prop:respectc}, the other facets $[P_F]$ of $[P]$ which are on the boundary of $[B_M(C)]^{\mathcal{D} \star}(C \cap D)$ are such that $[P_F] \subset [B_M(C)]^\star \subset [B_G]^{\star}$.
\end{itemize}
Therefore, the boundary of $[B_G]^{\mathcal{D} \star}(D)$ is either in faces of cubes which are in $\partial X$ or included in $[B_G]^{\star}$. 
\end{itemize}
Moreover, in each cube $C$ such that $D \cap C \neq \emptyset$ and $B_M(C) \neq \emptyset$, the boundary of $[B_M(C)]^{\mathcal{D} \star}(D) $ in $C - \partial C$ is $[B_M(C)]^{\mathcal{D} \star}(D) \cap ([B_M(C)]^\star - \partial [B_M(C)]^\star)$ (proposition \ref{prop:respectc}). Taking the union of these sets for all cubes of $G$, it can easily be seen that the boundary of $[B_G]^{\mathcal{D} \star}(D) $ in $X - \partial X$ is $[B_G]^{\mathcal{D} \star}(D) \cap ([B_G]^\star - \partial [B_G]^\star)$.
$\square$  %\vspace{0.5 cm}

\begin{proposition}
For all points $x \in X - [B_G]^\star$, there exists a unique set $D \in \mathcal{D}(B_G, G)$ such that $x \in [B_G]^{\mathcal{D} \star}(D)$.
\label{prop:cseparationG2}
\end{proposition}

\textbf{Proof.}
Let $C \in \mathcal{F}_d(G)$ be such that $x \in C$. 

If $B_M(C) = \emptyset$, then there exist a unique set $D \in \mathcal{D}(B_G, G)$ such that $C \subset D$.

Otherwise, there exists a unique set $D_C \in \mathcal{D}(B_M(C), C)$ such that $x \in [B_M(C)]^{\mathcal{D} \star}(D_C)$ (because of propositions \ref{prop:union} and \ref{prop:inter}) and there exists a unique set $D \in \mathcal{D}(B_G, G)$ such that $D_C \subset D$.
$\square$ \vspace{0.5 cm}

The classification function by the resistar approximation is defined directly from proposition \ref{prop:cseparationG2}.

\begin{definition}
The classification $[B_G]^{\star}(.)$ by the resistar approximation on the grid is a function from $X$ to $\{-1,0,+1\}$ defined for $x \in X$ as follows: 
\begin{itemize}
	\item If $x \in [B_G]^{\star}$, $[B_G]^{\star}(x) = 0$
	\item Otherwise, proposition \ref{prop:cseparationG2} ensures that there exists a unique classification set $ D \in \mathcal{D}(B_G, G)$ such that $x \in  [B_G]^{\mathcal{D} \star}(D)$, and $[B_G]^{\star}(x) = \mathcal{M}(v), v \in \mathcal{V}(D)$.
\end{itemize}
\end{definition}

This definition is consistent with the classification of the points of the grid by $\mathcal{M}$. Indeed, for all grid points $v \in \mathcal{V}(D)$, $D \in \mathcal{D}(B_G, G)$, $[B_G]^{\star}(v) = \mathcal{M}(v)$ and any point $x \in [B_G]^{\mathcal{D} \star}(D)$ is on the same side of $[B_G]^\star$ as $v$, because the boundary of $[B_G]^{\mathcal{D} \star}(D)$ in $X - \partial X$ is $[B_G]^{\mathcal{D} \star}(D) \cap ([B_G]^\star - \partial [B_G]^\star)$.

\begin{proposition}
For all $x \in X$, let $C_x$ be a cube of $\mathcal{F}_d(G)$ such that $x \in C_x$. We have:
	\begin{align}
	[B_G]^{\star}(x) = 	[B_M(C_x)]^\star(x).
\end{align}
\label{prop:classGrid}
\end{proposition}

\textbf{Proof.}
Consider $x \in X$ and $C_x \in \mathcal{F}_d(G)$ such that $x \in C_x$. 
There exists $D \in \mathcal{D}(B_G, G)$ such that $x \in  [B_G]^{\mathcal{D} \star}(D)$ (proposition \ref{prop:cseparationG2}). By definition, $[B_G]^{\mathcal{D} \star}(D) \cap C_x = [B_M(C_x)]^{\mathcal{D} \star}(D \cap C_x)$ thus $x \in [B_M(C_x)]^{\mathcal{D} \star}(D \cap C_x)$.
Therefore $[B_G]^{\star}(x) =	[B_M(C_x)]^\star(x)$.
$\square$ \vspace{0.5 cm}

Computing $[B_M(C)]^\star(x)$ can thus be performed by first computing a cube $C_x$ such that $x \in C_x$, and then applying algorithm \ref{Algo:cFaceProject} to $x$ in $C_x$. However, this approach would require to store the classification of all the vertices of the grid. The next subsection proposes a method requiring less memory space.

%------------------------------------------------
\subsection{Algorithm of classification avoiding to store the classification of all grid vertices} 
%------------------------------------------------
\label{sec:memory}
This subsection describes the algorithm of classification of c-resistar approximation when storing the boundary points and only the classification by $\mathcal{M}$ of the vertices of the edge on which the boundary point is located, instead of the classification by $\mathcal{M}$ of all the vertices of the grid. 

\subsubsection{Classification algorithm in the cube}
%------------------------------------------------

The modified classification algorithm is based on proposition \ref{th:vertexSign}. 

\begin{proposition} 
Let $C$ be a cube of $G$. For all $x \in C - [B]^\star$ and let $F_h$ be the face of $C$ such that $B_M(F_{h}) = \emptyset$ at the last step of algorithm \ref{Algo:cFaceProject} applied to $x$. The face $F_{h+1}$ from the previous step is such that $B_M(F_{h+1}) \neq \emptyset$ and there exists $b \in B_M(F_{h+1})$ such that $v_+(b) \in F_{h}$ or $v_-(b) \in F_{h}$.
\label{th:vertexSign}
\end{proposition}

\textbf{Proof}
The proof comes directly from the argument used in the proof of proposition \ref{prop:simplex}.
$\square$  \vspace{0.5 cm}

Algorithm \ref{Algo:classifc} modifies algorithm \ref{Algo:cFaceProject} using proposition \ref{th:vertexSign}: it tests the vertices of the boundary points in $B_{h+1}$ and the classification by $\mathcal{M}$ of the first of these vertices found in $F_h$ gives the final classification to return. Note that this algorithm supposes that $B_M(C) \neq \emptyset$ (which is not the case of algorithm \ref{Algo:cFaceProject}).

Moreover, computing the ray $\vec{r}(\bar{B}_M(F_i),x)$ requires to perform a division by $\left\|\bar{B}_M(F_i)  - x\right\|$ which may lead to very large numbers and strong losses of precision when $\left\|\bar{B}_M(F_i)  - x\right\|$ is very close to 0. Therefore, in practice, the classification returns $0$ when $\left\|\bar{B}_M(F_i)  - x\right\|$ is smaller than a given threshold (we take $0.00001$). With this modification, some points at small distance of $[B]^\star$ are classified $0$. Algorithm \ref{Algo:classifc} also includes this modification.

\begin{algorithm}[H]
%\dontprintsemicolon
\KwIn{$B$ non void set of boundary points in a cube $C \in \mathcal{F}_d(G)$, $x \in C$, $\delta$ small number}
	$F_d \leftarrow C$;
	$x_d \leftarrow x$;
	$B_d \leftarrow B$;
	$i \leftarrow d$;\\
	\While{$i \geq 0$}
	{\If{$\left\|x_{i} - \bar{B}_{i}\right\| < \delta$}{\Return 0}
	$x_{i-1} \leftarrow \vec{r}(\bar{B}_i , x_i) \cap \partial F_i$;\\
	$F_{i-1} \leftarrow F \in \mathcal{F}_{i-1}(F_i),  x_{i-1} \in F$;\\
	$B_{i-1} \leftarrow F_{i-1} \cap B_i$;\\
	\If{$B_{i-1} = \emptyset$}
	{\For{$b \in B_{i}$}
		{\If{$v_+(b) \in F_{i-1}$}{\Return $+1$}
		\If{$v_-(b) \in F_{i-1}$}{\Return $-1$}}}
	$i \leftarrow i-1$;}
\label{Algo:classifc}
\caption{c-resistar classification of point $x \in C$ with management of memory and numeric precision.}
\end{algorithm}

\subsubsection{Classification algorithm on the whole grid.}
%----------------------------------------

\begin{figure}
 	\centering
 			\includegraphics[width= 9 cm]{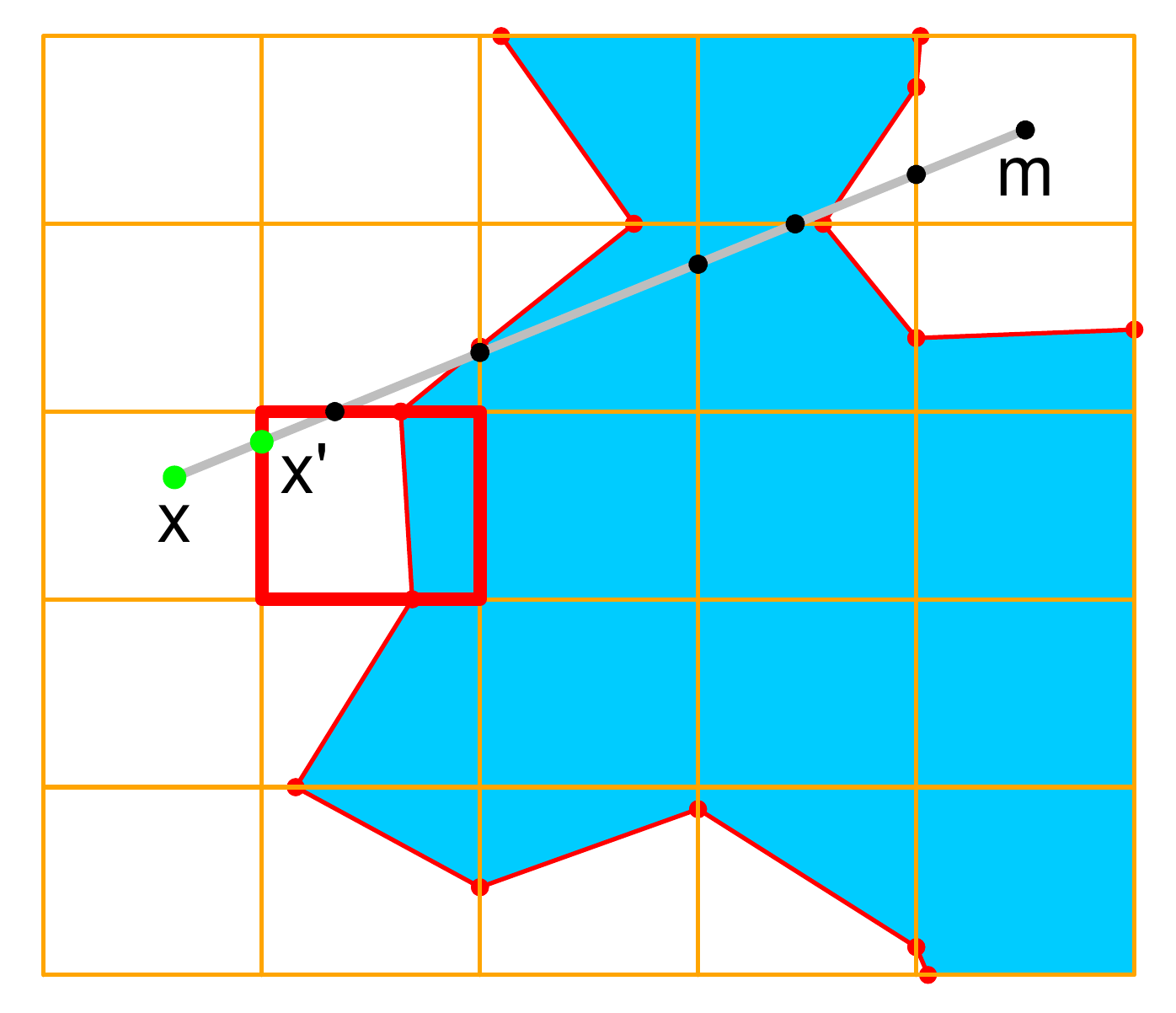} 
 	\caption{Example illustrating Algorithm \ref{Algo:resistarSurfaceClass} in 2 dimensions. Classifying point $x$ by the whole surface is performed by classifying point $x'$ by the resistar in the cube in bold, which is the first non-empty one among the ones cut by segment $[m,x]$, $m$ being the centre of a reference cube arbitrarily chosen.}
 	\label{fig:Classif3}
 \end{figure}

Algorithm \ref{Algo:resistarSurfaceClass} performs the classification of a point $x \in X$ by the c-resistar approximation on the grid keeping in memory only the set $C_G$ of cubes containing boundary points. It requires defining point $m \in X$ which the centre of an arbitrarily chosen cube $C_m$ in $C_G$. It computes the cube $C'_x \in C_G$ which is the closest to $x$ and such that $C'_x \cap [m,x] \neq \emptyset$ and the point $x' \in C'_x \cap [m,x]$ which is the nearest to $x$ in cube $C'_x$. Then, it computes the classification of $x'$ in the identified cube $C'_x$ (see illustration on Figure \ref{fig:Classif3}).

In practice, a hash table stores the set of couples $(C, B_M(C))$, for the cubes $C$ such that $B_M(C) \neq \emptyset$. Computing $x'$ and $C'_x$ can be done by computing the intersections of $[x, m]$ with the facets of cubes of the grid. The procedure then requires checking which cubes intersecting $[x, m]$ are in the hash table. The number of cubes crossing $[x, m]$ varies linearly with $n_G$ hence the number of requests to the hash table is also linear with $n_G$.  

\begin{algorithm}%[H]
%\dontprintsemicolon
\SetKwFunction{mResistarClass}{mResistarClass}
\SetKwFunction{keys}{keys}
\KwIn{$C_G = \{C \in \mathcal{F}_d(G), B_M(C) \neq \emptyset \}$, $m$ centre of a cube of $C_G$ chosen arbitrarily, $x$ point to classify}	
	%$\begin{aligned}
			%\hat{i} \leftarrow arg \min_{i = 1,...,k} \left\|\bar{B_i} - y\right\|;
	%\end{aligned}$\\
	%$m \leftarrow $ centre of $C_0$;\\
	$\mathscr{C} \leftarrow \left\{C \in C_G, C \cap [x,m] \neq \emptyset \right\}$ \\
	\For{$C \in \mathscr{C}$}
	{$\begin{aligned}
	x'(C) \leftarrow \mbox{arg } \min_{y \in [x, m] \cap C}\left\| x - y \right\|
	\end{aligned}$ }
	$\begin{aligned}
	C'_x \leftarrow \mbox{arg } \min_{C \in \mathscr{C}} \left\| x - x'(C) \right\| 
	\end{aligned}$ \\
%	$x' \leftarrow  $;\\
	\Return $[B_M(C'_x)]^\star(x'(C'_x))$
\label{Algo:resistarSurfaceClass}
\caption{Classification of a point by a resistar surface.}
\end{algorithm}

\begin{proposition}
When given a point $x \in X$ as input, algorithm \ref{Algo:resistarSurfaceClass} yields $[B_M(G)]^\star(x)$ as output.
\label{prop:classifcG}
\end{proposition}

\textbf{Proof.}
Let $x' = x'(C'_x)$ as defined in algorithm \ref{Algo:resistarSurfaceClass}. It always exists because $m$ is the centre of a cube $C_m$ containing boundary points, and $C'_x = C_m$ if no other cube containing boundary points cuts the segment $[x,m]$. By construction, the segment $[x, x']$ does not cross $[B_M(G)]^{\star}$, therefore $[B_M(G)]^{\star}(x) = [B_M(G)]^{\star}(x') = [B_M(C'_x)]^\star(x')$.
$\square$  %\vspace{0.5 cm}

%============================================================
\section{Manifold approximation with resistars in simplices from Kuhn triangulation.}
%============================================================

Figure \ref{fig:cresistar2D}, panel (b) and Figure \ref{fig:cresistar3D}, give examples of c-resistars that are not manifolds. Indeed, in both of these cases, it is impossible to find a continuous and invertible mapping from the neighbourhood of point $\bar{B }$ in the resistar into a hyperplane. However, for numerous applications, including viability kernel approximation, it is highly recommended to build manifold approximations.
In order to address this problem, we now break the cubes into simplices, using the Kuhn triangulation and we define resistars in these simplices.
 The next subsection focuses on the resistar in a single simplex and the following subsection on the approximation on the whole grid.

%---------------------------------------------------------------
\subsection{K-resistar in a single simplex from Kuhn triangulation.}
%---------------------------------------------------------------

%----------------------------------------------------------
\subsubsection{Definition and main properties.}
%----------------------------------------------------------
The Kuhn triangulation of a cube $C$ is defined by the set of $\mathcal{P}(d)$ permutations of set $\left\{1, 2, .., d\right\}$. Let $P$ be one of such permutations, the corresponding simplex $S_P$ is defined in cube $C$ as:
\begin{equation}
S_P = \left\{ x \in C, \min_{P(1)} \leq x_{P(1)} \leq x_{P(2)} \leq .. \leq x_{(P(d)} \leq \max_{P(d)} \right\}.
\end{equation}

Where $x_{P(i)}$ is the $P(i)^{th}$ coordinate of point $x$,  $\min_{P(1)}$ and $\max_{P(d)}$ are respectively the minimum value of the $P(1)^{th}$ coordinate and the maximum value of the $P(d)^{th}$ coordinate for points in cube $C$. 

\begin{figure}
 	\centering
 	\begin{tabular}{cc}
 		\includegraphics[width=6 cm]{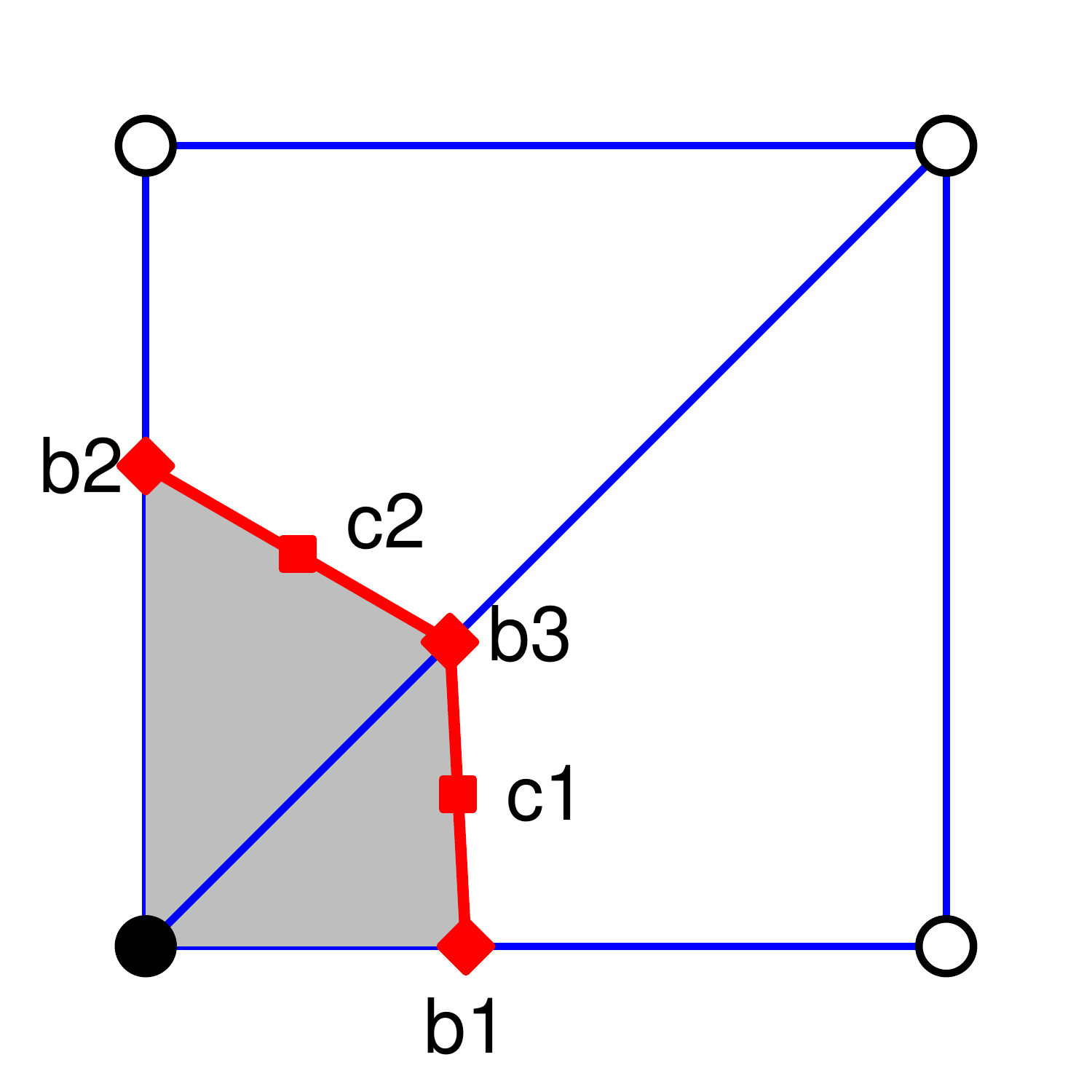} & 
 		\includegraphics[width= 6 cm]{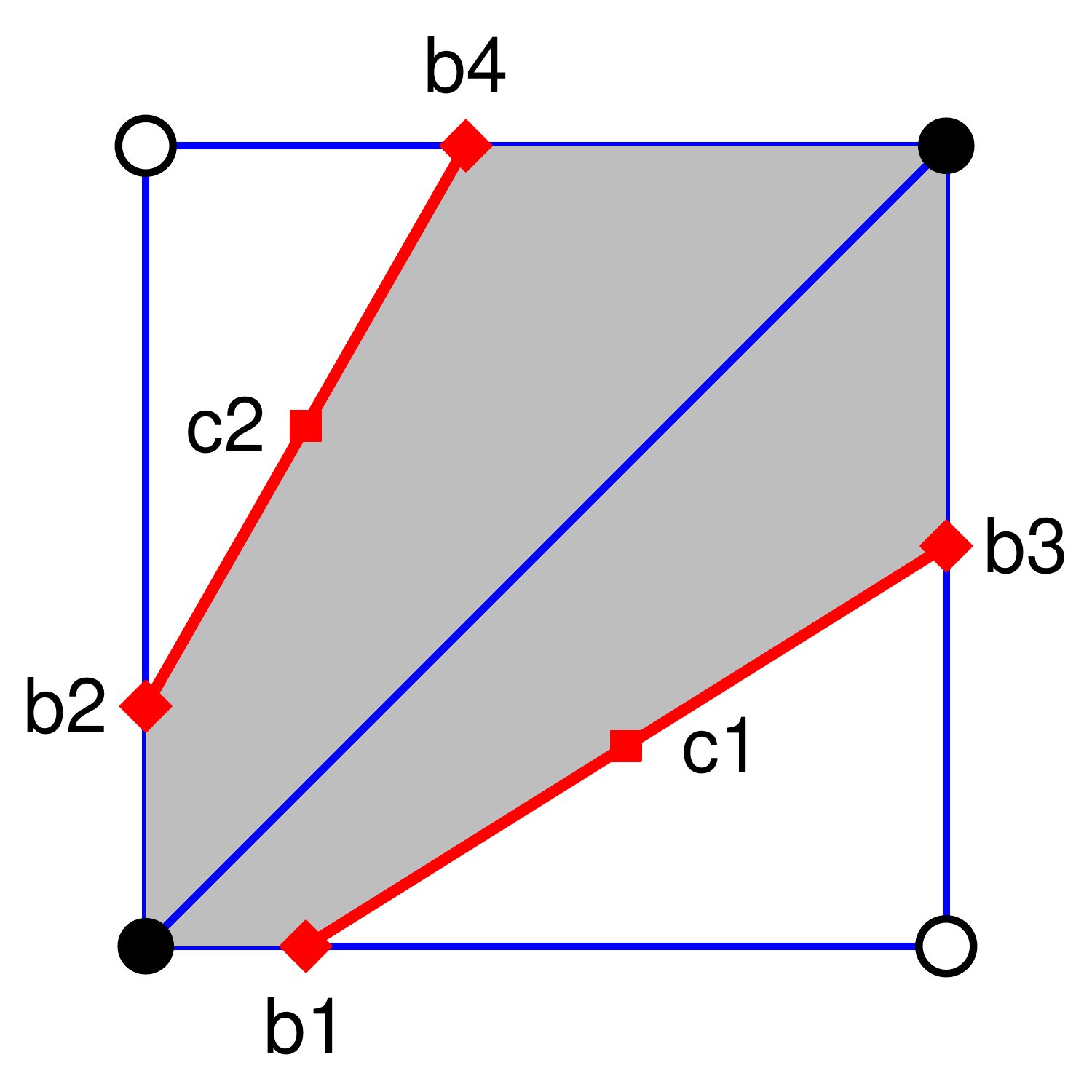} \\
		(a) & (b)
 	\end{tabular}
 	\caption{Examples of K-resistar in 2 D cubes. Panel (a): The cube includes two K-simplices with a resistar of two simplices in each. $[b_1, b_3]^\star =  [b_1, c_{1}] \cup [b_3, c_{1}] $ and $[b_2, b_3]^\star = [b_2, c_{2}] \cup [b_3, c_{2}] $ Panel (b): The cube includes two K-simplices, each including also a K-resistar of two simplices $[b_1, b_3]^\star = [b_1, c_{1}]$ $\cup$ $ [b_3, c_{1}] $ and $[b_2, b_4]^\star =  [b_2, c_{2}]$$\cup$ $ [b_4, c_{2}] $.}
 	\label{fig:kresistar2D}
 \end{figure}

\begin{figure}
 	\centering
 		\includegraphics[width=9 cm]{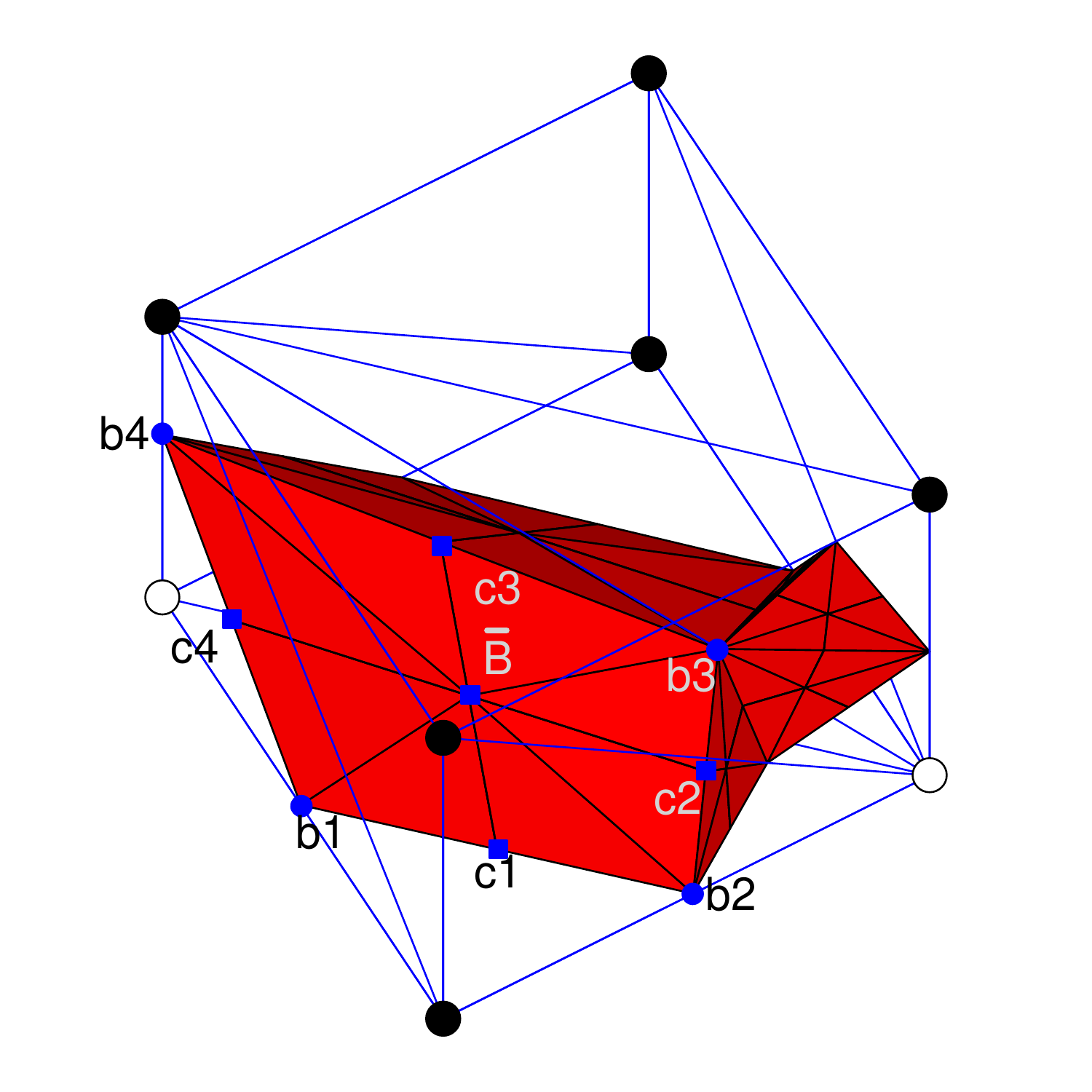} 
 	\caption{Example of K-resistars in K-simplices of a 3 D cube. The edges of the K-simplices are represented in blue. The boundary points on the cube edges are the same as on Figure \ref{fig:cresistar3D}. In total there are 6 K-resistars. For one of them, including 6 simplices, the vertices are represented. These vertices are among the four boundary points $B = \{b_1, b_2, b_3, b_4\}$, the four barycentre in 2D faces $\{ c_1, c_2, c_3, c_4 \}$, and the barycentre $\bar{B}$ of the four boundary points. More precisely, this K-resistar is: $[B]^\star =  [b_1, c_{1},\bar{B}]$ $\cup$ $[b_1, c_{4},\bar{B}]$ $\cup$ $[b_2, c_{1}, \bar{B}]$ $\cup$ $[b_2, c_{2}, \bar{B}]$ $\cup$ $[b_3,  c_{2}, \bar{B}]$ $\cup$ $[b_3,  c_{3}, \bar{B}]$ $\cup$ $[b_4, c_{3}, \bar{B}]$ $\cup$ $[b_4, c_{4}, \bar{B}]$.}
 	\label{fig:kresistar3D}
 \end{figure}

We denote these simplices with the prefix "K", in order to distinguish them from the simplices used to approximate $M$.  Examples of K-simplices in cubes are represented (by their edges) on Figures \ref{fig:kresistar2D} and \ref{fig:kresistar3D}. The set of $i$-dimensional faces of a K-simplex $S_P$ is denoted $\mathcal{K}_i(S_P)$. The set of K-simplices of cube $C$ is denoted $\mathcal{K}_d(C)$. The union of all the K-simplices defined by all the permutations is the cube itself:
\begin{equation}
C = \bigcup_{S_P \in \mathcal{K}_d(C)} S_P.
\end{equation}

On each edge  $[v_-,v_+]$ of a K-simplex such that $\mathcal{M}(v_-) = -1$ and $\mathcal{M}(v_+) = +1$, we compute a boundary point $b_M([v_-,v_+])$, approximating the intersection between the edge and $M$, by $q$ successive dichotomies, as previously. The value of $q$ may be adjusted in order to ensure a given precision of the approximation even on the longest edge of the K-simplex. 

We denote $B_M(S)$ the set of boundary points of a K-simplex or of a K-simplex face $S$. 

The resistars in K-simplices are defined similarly to the c-resistars, except that faces of the K-simplices are considered instead of faces of cubes (see examples on Figures \ref{fig:kresistar2D} and \ref{fig:kresistar3D}). We denote K-resistars the resistars defined in K-simplices.

\begin{definition}
Let $S_P$ be a K-simplex such that $B = B_M(S_P) \neq \emptyset$. The K-resistar $[B]^\star$ is the following set:
	\begin{align}
	 [B]^\star = \bigcup_{\{ F_1,...,F_{d-1} \} \in \mathcal{K}^\star(B,S_P)} [\{ \bar{B}_M(F_1),...,\bar{B}_M(F_{d-1}) \} \cup \{ \bar{B} \}]
	\end{align}
	with:
	\begin{align}
	\{ F_1,...,F_{d-1} \} \in \mathcal{K}^\star(B,S_P) \Leftrightarrow
		\begin{cases}
		F_1 \in \mathcal{K}_1(S_P), B_M(F_1) \neq \emptyset\\
		F_i \in \mathcal{K}_i(S_P), F_{i-1} \subset F_{i}, 2 \leq i \leq d-1.
		\end{cases}
\end{align}
where $i$ is an integer.
\end{definition}

Using the same arguments as in the proofs of propositions \ref{prop:simplex} and \ref{prop:border}, it can easily be shown that K-resistar $[B_M(S_P)]^\star$ is a set of $d-1$-dimensional simplices and that its boundary is included in the boundary of $S_P$. %Moreover, as stated in proposition \ref{prop:manifold}, a resistar in a K-simplex is always a manifold.

\begin{proposition}
Let $S_P$ be a K-simplex such that $B = B_M(S_P) \neq \emptyset$. $[B]^\star$ is a manifold. 
\label{prop:manifold}
\end{proposition}

\textbf{Proof.}
Consider a point $x \in [B]^\star$. There exists $\{F_1,...,F_{d-1} \} \in \mathcal{K}^\star(B, S_P)$ such that $x \in [\{ \bar{B}_M(F_1),..., \bar{B}_M(F_{d-1})\} \cup \{ \bar{B} \}]$ and there exist positive numbers $(\lambda_i), i \in  \left\{1,...,d\right\}$ such that (setting $F_d = S_P$):

	\begin{align}
	x = \sum_{i \in \left\{1,...,d\right\}} \lambda_i \bar{B}_M(F_i) \mbox{   and } \sum_{i \in \left\{1,...,d\right\}} \lambda_i = 1.
	\label{eq:xcombFm}
	\end{align}
	
The Vapnik-Chervonenkis dimension of the $d-1$-dimensional linear separators being $d+1$ \cite{Vapnik1971}, for any cut into two sets of $d+1$ affinely independent points, there exists a $d-1$-dimensional hyperplane making this cut.
Let $H$ be a hyperplane separating the set $V_+$ of vertices $v$ of $S_P$ such that $\mathcal{M}(v) = +1$ from the set $V_-$ of vertices $v$ of $S_P$ such that $\mathcal{M}(v) = -1$. To each boundary point $b \in B$, located on edge $[v_-, v_+]$ with $v_- \in V_-$ and $v_+ \in V_+$, we associate point $b_H = [v_-, v_+] \cap H$. Let $B_H$ be the set of points $b_H$ so defined. For any face $F_i$ of $S_P$, we denote $B_H(F_i) = F_i \cap B_H$.
We can associate uniquely to $x$ the point $x_H \in H$ defined as follows:
	
	\begin{align}
	x_H = \sum_{i \in \left\{1,...,d\right\}} \lambda_i \bar{ B}_H (F_i),
	\end{align}
	
	where the faces $F_i$ and $\lambda_i$ are defined in equation \ref{eq:xcombFm}. Therefore there exists a continuous and invertible function from $[B]^\star$ to hyperplane $H$.	 
$\square$ % \vspace{0.5 cm}

\subsubsection{Classification function and algorithm.}
\begin{figure}
 	\centering
 	\begin{tabular}{cc}
 		\includegraphics[width=6 cm]{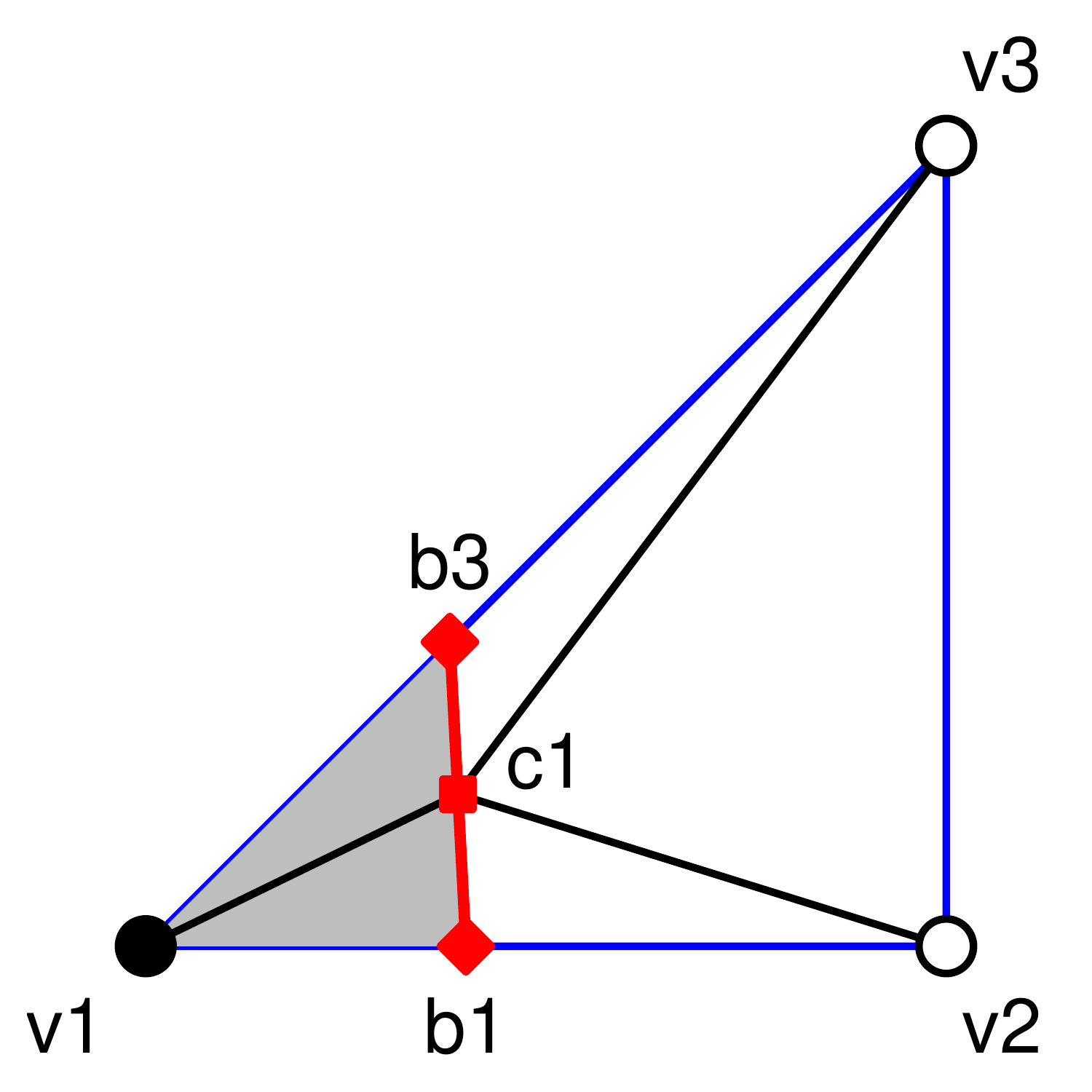} & 
 		\includegraphics[width= 6 cm]{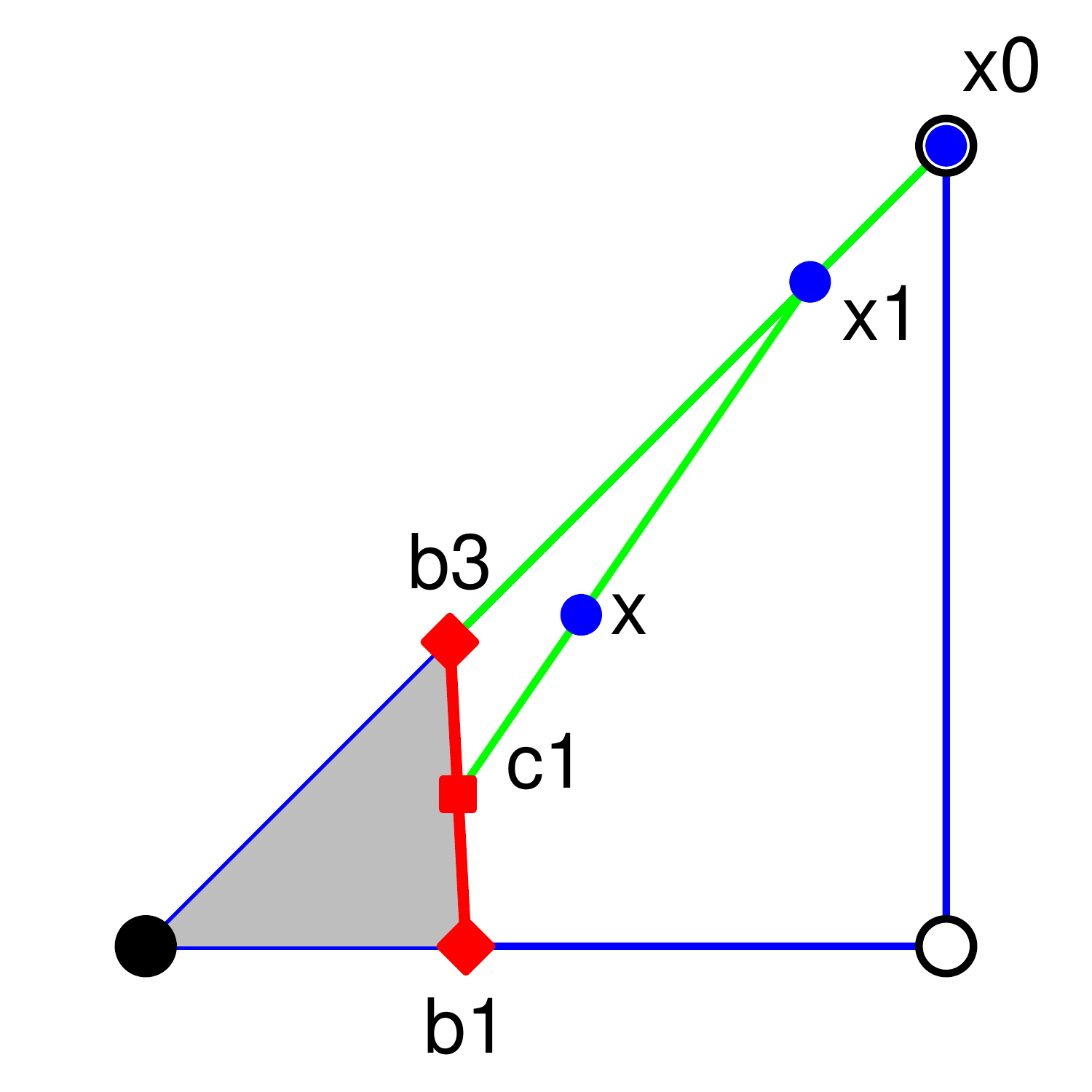} \\
		(a) & (b)
 	\end{tabular}
 	\caption{Panel (a): Examples of faces without boundary points $[V_\pm]$ and corresponding classification sets $[B]^{\mathcal{D} \star}([V_\pm])$ of a 2D K-resistar. The vertices of the K-simplex in black are classified -1 by $\mathcal{M}(.)$ and ones in white are classified +1. $[V_-] = \{ v_1,\}$ and $[V_+] = [v_2, v_3] $ and the classification sets are: $[B]^{\mathcal{D} \star}([V_-]) = [v_1, b_3, c_1] $$\cup [v_1, b_1, c_1]$ and $[B]^{\mathcal{D} \star}([V_+]) = [v_2, b_1, c_1]$$ \cup [v_2, v_3, c_1]$$ \cup [v_3, b_3, c_1]$ . Panel (b): Illustration of the classification algorithm. $x_1 = \vec{r}(c_1,x) \cap \partial S_P$ and $x_1$ belongs to face $F$ which includes boundary point $b_3$. $ x_0 = \vec{r}(b_3,y_1) \cap \partial F$. $x_0$ is a vertex such that $\mathcal{M}(x_0) = +1$, therefore $[b_1, b_3]^\star(x) = +1$.}
 	\label{fig:kresistar2DCClassif}
 \end{figure}

%-------------------------------------------
As stated in details in proposition \ref{prop:Kseparation}, a K-simplex $S_P$ such that $B_M(S_P) \neq \emptyset$ is always separated by its K-resistar into two classification sets, each containing a single face of the K-simplex without boundary points (Figure \ref{fig:kresistar2DCClassif} panel (a) shows examples of these classification sets in 2 dimensions).

\begin{proposition}
Let $S_P$ be a K-simplex such that $B = B_M(S_P) \neq \emptyset$. let $\mathcal{H}(B,S_P)$ be the set of faces of $S_P$ without boundary points :
\begin{align}
	\mathcal{H}(B,S_P) = \left\{ F \in \mathcal{K}(S_P), B_M(F) = \emptyset \right\},
\end{align}
where $\mathcal{K}(S_P)$ is the set of all the faces of $S_P$. $\mathcal{H}(B,S_P)$, includes 2 faces of $S_P$: $[V_+]$ where $V_+ = \{ v \in \mathcal{V}(S_P), \mathcal{M}(v) = +1 \}$ and $[V_-]$ where $V_- = \{ v \in \mathcal{V}(S_P), \mathcal{M}(v) = -1 \}$. The resistar $[B]^\star$ separates $S_P$ into two classification sets $[B]^{\mathcal{D} \star}([V_+])$ and $[B]^{\mathcal{D} \star}([V_-])$, such that $[V_\pm] \subset [B]^{\mathcal{D} \star}([V_\pm])$, which are :
\begin{align}
	[B]^{\mathcal{D} \star}([V_\pm]) = \bigcup_{\{ F_h,..,F_{d-1} \} \in \mathcal{K}^{\mathcal{D} \star}([V_\pm],S_P)} [\mathcal{V}(F_h) \cup \{\bar{B}(F_1),...,\bar{B}_M(F_{d-1}) \} \cup \{ \bar{B} \}],
	\end{align}
with:
	\begin{align}
	\{ F_h,..,F_{d-1} \} \in \mathcal{K}^{\mathcal{D} \star}([V_\pm], S_P) \Leftrightarrow
	\begin{cases}
	F_h  \in \mathcal{K}_h([V_\pm]), 0 \leq h \leq dim([V_\pm])\\
	F_i \in \mathcal{K}_i(S_P), F_{i-1} \subset F_i, h+1 \leq i \leq d-1, \\
	B_M(F_{h+1}) \neq \emptyset, 
	\end{cases}
\end{align}
where dim$([V_\pm])$ is the dimensionality of $[V_\pm]$. 

Moreover, $[B]^{\mathcal{D} \star}([V_+])$ and $[B]^{\mathcal{D} \star}([V_-])$ are connected sets, $[B]^{\mathcal{D} \star}([V_-])\cap [B]^{\mathcal{D} \star}([V_+]) = [B]^\star$ and $[B]^{\mathcal{D} \star}([V_-]) \cup [B]^{\mathcal{D} \star}([V_+]) = S_P$.
\label{prop:Kseparation}
\end{proposition}

\textbf{Proof.}
The sets $V_+ $ and $V_- $ are such that the sets $[V_+]$ and $[V_-]$ are faces of $S_P$, because $S_P$ is a simplex and any set of its vertices defines one of its faces. %Therefore $\mathcal{H}(B,S_P)$ comprises two faces of $S_P$: $[V_+]$ and $[V_-]$.

The polytopes of $[B]^{\mathcal{D} \star}([V_\pm])$ include at least one face $F_h$ of $[V_\pm]$, and $[V_\pm] \subset [B]^{\mathcal{D} \star}([V_\pm])$ and is a connected set, therefore $[B]^{\mathcal{D} \star}([V_\pm])$ is a connected set.

Using the same reasoning as in the proof of proposition \ref{prop:respectc}, it can easily be shown that $[B]^{\mathcal{D} \star}([V_-]) \cap [B]^{\mathcal{D} \star}([V_+]) = [B]^\star$ and $[B]^{\mathcal{D} \star}([V_-]) \cup [B]^{\mathcal{D} \star}([V_+]) =S_P$.
$\square$  \vspace{0.5 cm}

The following definition of the classification function is consistent because proposition \ref{prop:Kseparation} ensures that, if $x \notin [B]^\star$, either $x \in [B]^{\mathcal{D} \star}([V_+])$ or $x \in [B]^{\mathcal{D} \star}([V_-])$.

\begin{definition}
The classification $[B]^\star(.)$ by resistar $[B]^\star$ in K-simplex $S_P$ is the function from $S_P$ to $\left\{-1, 0, +1\right\}$ which, is defined as follows for $x \in S_P$ (setting $B = B_M(S_P)$):
\begin{itemize}
\item If $B = \emptyset$, then $[B]^\star(x) = \mathcal{M}(v)$, $v \in \mathcal{V}(S_P)$;
\item Otherwise:
\begin{itemize}
	\item If $x \in [B]^\star$, $[B]^\star(x) = 0$;
	\item Otherwise, let $\mathcal{H}(B,S_P) = \{[V_-], [V_+] \}$:	
	\begin{itemize}
		\item If  $x \in [B]^{\mathcal{D} \star}([V_+])$, $[B]^\star(x) = +1$;
		\item If $x \in [B]^{\mathcal{D} \star}([V_-])$, $[B]^\star(x) = -1$.
	\end{itemize}
\end{itemize}
\end{itemize}
\end{definition}

Algorithm \ref{Algo:cFaceProject} is easily adapted to a K-simplex instead of a cube, by replacing the faces of the cube by the faces of the K-simplex (see Figure \ref{fig:kresistar2DCClassif} panel (b)). It is direct to show that this algorithm returns $[B]^\star(x)$.

%----------------------------------------------------
\subsection{K-resistar approximation on the grid.} 
%----------------------------------------------------

\subsubsection{Definition.}
%----------------------------------------

The definition of the K-resistar approximation of $M$ on the grid is similar to one of the c-resistar approximation. 
 
\begin{definition}
Let $B_M^K(G)$ be the set of all the boundary points of all the edges of the K-simplices of the grid. We call K-resistar approximation of $M$ on grid $G$, denoted $[B_M^K(G)]^{\star}$, the following set of simplices:

	\begin{align}
	 [B_M^K(G)]^{\star}= \bigcup_{S_P \in \mathcal{K}_d(G), B_M(S_P) \neq \emptyset} [B_M(S_P)]^{\star}, 
\end{align}
where $\mathcal{K}_d(G)$ is the set of all the K-simplices defined in the cubes of $G$.
\end{definition}

\begin{proposition}
The K-resistar approximation of $M$ on grid $G$ is a set of $d-1$-dimensional simplices and its boundary is included in the boundary of $X$.
\label{prop:borderKg}
\end{proposition}

\textbf{Proof.}
The proof is the same as the one of proposition \ref{prop:bordercg}, considering faces of K-simplices instead of cube faces.
$\square$ % \vspace{0.5 cm}

\begin{proposition}
The K-resistar approximation of $M$ on grid $G$ is a $d-1$-dimensional manifold.
\end{proposition}

\textbf{Proof.}
K-resistars in K-simplices are $d-1$-dimensional manifolds as shown in proposition \ref{prop:manifold}. We need to show that the union of K-resistars in K-simplices sharing a K-simplex face is also a manifold in the neighbourhood of their common points belonging to this face.

Let $F_i$ be a $i$-dimensional face of a K-simplex, such that $B_M(F_i) \neq \emptyset$. $[B_M(F_i)]^\star = [B_M^K(G)]^\star \cap F_i$ is a $i-1$-dimensional manifold and there exists a $i-1$-dimensional hyperplane $H_i$ which separates the vertices of $F_i$ classified positively by $\mathcal{M}$ from the ones classified negatively. Let $n_i$ be the normal vector of $H_i$, such that the vertices classified positively by $\mathcal{M}$ are on the positive side of $H_i$. % Let $\sigma_h$ be the set of K-simplices of $C$ that share the face $F_h$. %For any point $x \in [B \cap F_h]^\star$, there exists a neighbourhood $U$ of $x$ in $C$ which is included in the union of the K-simplices sharing $F_h$ because the union of the K-simplices is $C$. 

For each K-simplex $S$ in $\mathcal{K}_d(G)$ containing $F_i$, there exists a $d-1$-dimensional hyperplane $H_S$ separating the vertices of $S$ classified positively by $\mathcal{M}$ from the ones classified negatively. Let $n_S$ be the normal vector of  $H_S$, such that the vertices classified positively by $\mathcal{M}$ are on the positive side of $\mathcal{H_S}$. 

We have $n_i.n_S \geq 0$ because the hyperplane $H_S$ also separates the positive and negative vertices of $F_i$. Therefore, we can choose each $H_S$ such that $H_S \cap F_i = H_i$.  With this choice of the hyperplanes $H_S$, the continuous and bijective mapping between each simplex of the K-resistar in $S$ and $ H_S$, defined in the proof of proposition \ref{prop:manifold}, is the same for the points in $[B_M(F_i)]^\star$ for all K-simplices $S$ sharing face $F_i$. 

Let $H'_i$, be the $d-1$ dimensional hyperplane extending $H_i$ to the set $X$ by extending the $i$-dimensional normal vector $n_i$ to the $d$-dimensional normal vector $n'_i$ of $d-1$-dimensional hyperplane $H'_i$ which coincides with $H_i$ in $F_i$, by setting to $0$  all the coordinates of $n'_i$ which are not defined in $F_i$. The composition of the mappings from the K-resistar in $S$ to the hyperplane $H_S$ with the orthogonal projection on $H'_i$ for all $S$ defines a continuous and bijective mapping from the neighbourhood of $[B_M(F_i)]^\star$ in the K-resistar approximation to hyperplane $H'_i$.
$\square$  %\vspace{0.5 cm}

\subsubsection{Classification by the K-resistar approximation on the grid.}
%----------------------------------------

The classification sets in $X$ defined by the K-resistar approximation on the grid $G$ are defined similarly to those of the c-rersistar approximation on the grid. $\mathcal{K}_i(G)$ denotes the set of all the faces of K-simplices defined in the cubes of the grid ($\mathcal{K}_d(G)$ denoting the set of all the K-simplices defined in the grid cubes).

\begin{definition}
Let $B_G = B_M^K(G)$ be the set of boundary points defined in all the K-simplices of grid $G$. Let $\mathcal{H}^K(B_G,G) = \left\{ F \in \mathcal{K}(G), F \cap B_G = \emptyset \right\}$, where $\mathcal{K}(G)$ is the set of the K-simplices of the cubes of the grid $G$ and of all their faces. Let $\mathcal{D}^K(B_G,G)$ be the connected components of $\mathcal{H}^K(B_G,G)$. 

For $D \in \mathcal{D}^K(B_G,G)$, the classification sets $[B_G]^{\mathcal{D}\star}(D)$ defined by the K-resistar approximation are:
\begin{align}
	[B_G]^{\mathcal{D}\star}(D) = \bigcup_{ S_P \in \mathcal{K}_d(G) } [B_M(S_P)]^{\mathcal{D} \star}(D \cap S_P), 
\end{align}
with:
\begin{align}
[B_M(S_P)]^{\mathcal{D} \star}(D \cap S_P) =
	\begin{cases}
	\emptyset, \mbox{ if } D \cap S_P = \emptyset,\\
	S_P, \mbox{ if } D \cap S_P = S_P,
	\end{cases}
\end{align}
and is specified in proposition \ref{prop:Kseparation} in the other cases. 
\end{definition}

\begin{proposition}
 For all $D \in \mathcal{D}^K(B_G,G)$, $D \subset [B_G]^{\mathcal{D}\star}(D)$, $[B_G]^{\mathcal{D}\star}(D)$ is connected and its boundary in $X - \partial X$ is a subset of $[B_G]^{\star}$.
\label{prop:KseparationG}
\end{proposition}

\textbf{Proof.}
The proof is similar to the one of proposition \ref{prop:cseparationG}, when using faces of K-simplices instead of cube faces.
$\square$ % \vspace{0.5 cm}

\begin{proposition}
For all points $x \in X$, such that $x \notin [B_G]^\star$, there exists a unique set $D \in \mathcal{D}^K(B_G, G)$ such that $x \in [B_G]^{\mathcal{D} \star}(D)$.
\label{prop:KseparationG2}
\end{proposition}

\textbf{Proof.}
The proof is similar to the one of proposition \ref{prop:cseparationG2}.
$\square$ 

\begin{definition}
Let $B_G = B_M^K(G)$. The classification $[B_G]^{\star}(.)$ by the resistar approximation from the grid is the function from $X$ to $\{-1,0,+1\}$ defined as follows for $x \in X$: 
\begin{itemize}
	\item If $x \in [B_G]^{\star}$, $[B_G]^{\star}(x) = 0$
	\item Otherwise, proposition \ref{prop:KseparationG} ensures that there exists a unique set $D \in \mathcal{D}^K(B_G,G)$ such that $x \in [B_G]^{\mathcal{D}\star}(D)$, and $[B_G]^{\star}(x) = \mathcal{M}(v), v \in \mathcal{V}(D)$.
\end{itemize}
\end{definition}

\begin{proposition}
For $x \in X$, let $S_P(x)$ be the K-simplex of $\mathcal{K}_d(G)$ such that $x \in S_P(x)$. We have:
	\begin{align}
	[B_M^K(G)]^{\star}(x) = 	[B_M(S_P(x))]^\star(x).
\end{align}
\end{proposition}

\textbf{Proof.}
The proof is similar to the one of proposition \ref{prop:classGrid}.
$\square$

\subsubsection{Classification algorithm.}
%----------------------------------------

In practice, when classifying point $x \in X$, we also use algorithm \ref{Algo:resistarSurfaceClass}, which provides a cube $C'_x$ such that $B_M^K(C'_x) \neq \emptyset$ and a point $x'$ to classify in this cube, ensuring $[B_M^K(G)]^\star(x) = [B_M^K(C'_x)]^\star(x')$. Then a K-simplex of $C'_x$ containing $x'$ is determined by ordering the coordinates of $x'$, this order providing the permutation $P$ defining $S_P$. Then the procedure computes $B_M^K(C'_x) \cap S_P$, the boundary points belonging to $S_P$. At this point, the classification of $x'$ in K-simplex $S_P$ is performed with algorithm \ref{Algo:classifc} (in its version adapted to K-simplices). 

The procedure of classification using K-resistars is thus a bit more complicated than the one of c-resistars because it requires identifying the K-simplex containing the point to classify and determining its boundary points.

%==========================================
\section{Accuracy of the approximation when the size of the grid increases.}
%===========================================

Theorem \ref{th:accuracy1} bounds the Hausdorff distance between a resistar approximation and the manifold to approximate, when this manifold is smooth enough. 

The distance $d(x, A)$ from a point $x \in \mathbb{R}^d$ to a set $A \subset \mathbb{R}^d$ is defined as ($\inf$ denoting the infimum):
\begin{align}
	d(x, A) = \inf_{y \in A} \left\|x - y\right\|.
\end{align}
The Hausdorff distance $d_H(A,B)$ between set $A$ and set $B$, both subsets of $\mathbb{R}^d$, is defined as  ($\sup$ denoting the supremum):
\begin{align}
	d_H(A, B) = \max \left(\sup_{x \in A} d(x, B), \sup_{y \in B} d(y, A)\right).
\end{align}
The smoothness of the manifold is characterised by its \emph{reach} \cite{Federer1959}, which is the supremum of $\rho$ such that for any point $x$ of $\mathbb{R}^d$ for which $d(x, M) = \rho$, there is only one point $y \in M$ such that $\left\|x - y \right\| = \rho$. Note that if the reach of $M$ is strictly positive, then $M$ is twice differentiable \cite{Federer1959}.

\begin{theorem}
Let $M$ be a $d-1$-dimensional manifold cutting the compact $X = [0,1]^d$ into two parts, $G$ be a regular grid of $n_G^d$ points covering $X$ and its boundary, $\epsilon = \frac{1}{n_G-1}$ be the size of an edge of the grid.

If the reach $r$ of $M$ is such that $r > \sqrt{2} d \epsilon $, if for all $i$-dimensional faces $F$ of $X$, $M \cap F$ is a $i-1$-dimensional manifold of reach $r_F > \sqrt{2} i \epsilon$, and if all the boundary points are determined with $q \geq -\log_2(\epsilon)$ dichotomies, then the Hausdorff distance between $M$ and its resistar approximation (in cubes or in K-simplices) decreases like $\mathcal{O}(d \epsilon^{2})$.
 \label{th:accuracy1}
\end{theorem} 

The proof of theorem \ref{th:accuracy1} uses two lemmas presented in paragraph \ref{sec:lemmas}. Then it uses an induction on the space dimensionality (set in paragraph \ref{sec:induc0}) with two parts: bounding the distance between the resistar approximation and $M$ (paragraph \ref{sec:induc1}) and bounding the distance between $M$ and the resistar approximation (paragraph \ref{sec:induc2}). 

%----------------------------------------
\subsection{Lemmas.}
%----------------------------------------
\label{sec:lemmas}
\begin{lemma}
Let $r >0$ be the reach of $M$ and let $\delta > 0$ be such that $r > \delta$. For any couple of points $(y_1,y_2)$ of $M$ such that $\left\|y_1 - y_2 \right\| \leq \delta$:
\begin{equation}
 \left\|y_2 - P_{y_1}(y_2) \right\| \leq \frac{\delta^2}{2r} + \mathcal{O}\left( \delta^3\right),
\end{equation}
where $P_{y_1}(y_2)$ is the orthogonal projection of $y_2$ on the hyperplane tangent to $M$ at $y_1$.  
\label{th:distTangent}
\end{lemma}

\textbf{Proof.}
Let $(y_1,y_2) \in  M^2$ be such that $\left\|y_1 - y_2 \right\| \leq \delta < r$, and $n$ be the normal vector of the hyperplane $T$ tangent to $M$ at $y_1$. Let $M_-$ (resp. $M_+$) be the set of points $x \in X$ such that $\mathcal{M}(x) = -1$ (resp.  $\mathcal{M}(x) = +1$). There exists $\mathcal{B}(c_+,r) $ (resp.  $\mathcal{B}(c_-,r) $) a ball tangent to $M$ at $y_1$ such that $M_- \cap \mathcal{B}(c_+,r) = \emptyset$ (resp. $M_+ \cap \mathcal{B}(c_-,r) = \emptyset$). $y_2$ is located between the balls $\mathcal{B}(c_+,r) $ and $\mathcal{B}(c_-,r) $, and we can suppose that it is closer to $\mathcal{B}(c_+,r) $ (the reasoning would of course be the same if it was closer to  $\mathcal{B}(c_-,r) $). Let $y'_2$ be the projection of $y_2$ on $\partial \mathcal{B}(c_+,r)$ parallel to $n$. We have:
\begin{align}
\left\|y_2 - P_{y_1}(y_2) \right\| \leq \left\|y'_2 - P_{y_1}(y_2) \right\|,
\end{align}
and:
\begin{align}
\left\|P_{y_1}(y_2) - y_1 \right\| \leq \delta,
\end{align}
because $P_{y_1}(y_1) = y_1$ and the projection is contracting. 
Moreover (see figure \ref{fig:accuracy1}, panel (a)):  
\begin{align}
\left\|y'_2 - P_{y_1}(y_2) \right\| = r (1 - \cos \alpha),
\label{eq:projdist}
\end{align}
with:
\begin{align}
\cos \alpha \geq \sqrt{1 - \left(\frac{\delta}{r}\right)^2},
\end{align}
Developing equation \ref{eq:projdist} at the second order, we get:
\begin{equation}
 \left\|y_2 - P_{y_1}(y_2) \right\| \leq \frac{\delta^2}{2r} + \mathcal{O}(\delta^{3}).
	\label{eq:accuracy1}
\end{equation}
$\square$  %\vspace{0.5 cm}

\begin{lemma}
Let $\mathscr{C}$ be a set of $j^d$ adjacent cubes of the grid, covering a cubic part of $X$ of edge size $j \epsilon$ ($\epsilon = \frac{1}{n_G -1}$), including a non-void set of boundary points $B = B_M(\mathscr{C})$, computed on cube edges, or $B = B_M^K(\mathscr{C})$, computed on edges of K-simplices, each determined with $q$ dichotomies. If the reach $r$ of $M$ is such that:
\begin{align}
 r > j\sqrt{d} \epsilon, 
\end{align}
then for any point $x \in [B]^\star$ and for any point $y \in (M \cap \mathscr{C})$:
\begin{equation}
 \left\|x - P_y(x) \right\| \leq  2^{-q}  \epsilon + \frac{d j^2}{2r} \epsilon^2 + \mathcal{O}(\epsilon^{3}),
\end{equation}
where $P_y(x)$ is the orthogonal projection of $x$ on the hyperplane tangent to $M$ at $y$.  
\label{th:projTangent}
\end{lemma}

\textbf{Proof.}
By construction, for any boundary point $b$ of $B$ located on edge $[v_-(b), v_+(b)]$ there exists a point $b_M \in M \cap [v_-(b), v_+(b)]$ such that $\left\|b - b_M \right\| \leq 2^{-q-1} \epsilon$. Therefore, for each boundary point $b$ of $B$, we can write:

\begin{equation}
	\left\|b  - P_y(b) \right\| \leq \left\|b - b_M  \right\|+ \left\|b_M - P_y(b_M) \right\|+ \left\|P_y(b_M) - P_y(b)\right\|,
	\end{equation}
	
with:
\begin{itemize}
\item $\left\|b - b_M\right\| \leq 2^{-q-1} \epsilon$ (see equation \ref{eq:precision});
\item $\left\|P_y(b_M) - P_y(b)\right\| \leq  2^{-q-1} \epsilon$, because the orthogonal projection is contracting;
\item  $\left\|b_M - P_y(b_M) \right\| \leq \frac{d j^2 \epsilon ^2}{2r} + \mathcal{O}(\epsilon^{3})$, because $\left\|b_M - y\right\| \leq j\sqrt{d}\epsilon$, since both $b_M$ and $y$ belong to $\mathscr{C}$, and applying lemma \ref{th:distTangent}. 
\end{itemize}
Overall, we get:

\begin{equation}
	\left\|b  - P_y(b) \right\| \leq 2^{-q} \epsilon + \frac{dj^2 }{2r} \epsilon^2 + \mathcal{O}(\epsilon^{3}).
	\end{equation}
	
Moreover, by definition of $[B]^\star$, for any point $x$ in $[B]^{\star}$ there exists a set of positive numbers $(\lambda_b)_{b \in B}$ such that:

\begin{equation}
	x = \sum_{b \in B} \lambda_b b, \mbox{  with: } \sum_{b \in B} \lambda_b = 1.
	\label{eq:accuracy2}
\end{equation}	
	
Therefore:
\begin{equation}
 \left\|x - P_y(x)\right\| \leq \sum_{b \in B} \lambda_b \left\| b - P_y(b) \right\| \leq 2^{-q} \epsilon + \frac{d j^2 }{2r} \epsilon^2 + \mathcal{O}(\epsilon^{3}). 
\end{equation}	
$\square$  %\vspace{0.5 cm}

\begin{figure}
 	\centering
 	\begin{tabular}{cc}
 		\fbox{\includegraphics[width=6.5 cm]{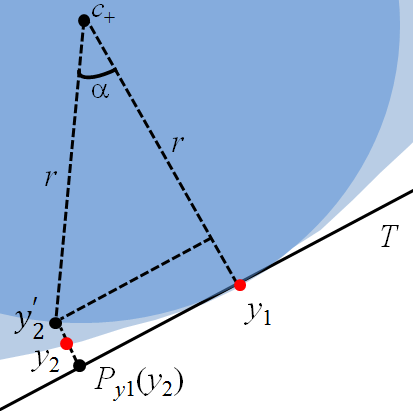}} & 	\fbox{\includegraphics[width= 6.5 cm]{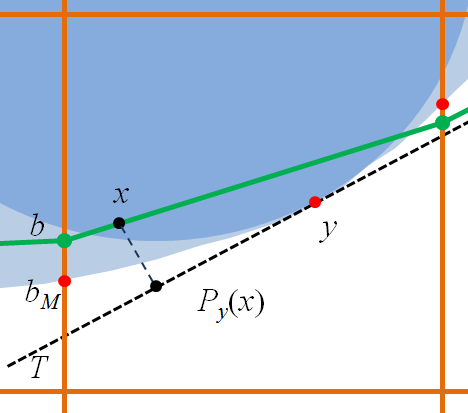}} \\
		(a) & (b)
 	\end{tabular}
 	\caption{Panel (a): Illustration of lemma \ref{th:distTangent}. Computing distance $\left\|y'_2 - P_{y_1}(y_2)\right\|$ uses angle $\alpha$. Panel (b):  Illustration of lemma \ref{th:projTangent}. The resistar is represented in green and the limits of grid cubes in orange. The distance from $x$ located on the resistar to the tangent hyperplane $T$ is bounded using the distance between the boundary points and $T$. In both panels, the region in light blue represents the positive region $M_+$ defined by $M$ and the region in darker blue is the positive tangent ball.}
 	\label{fig:lemmas}
 \end{figure} 
 
%\begin{theorem}
%If all the boundary points $B_f$ of the resistars are determined with at least $\log_2(n_G)$ dichotomies and for any point $y$ of $\partial f \cap [0,1]^d$, and the reach $r$ of $\partial f$ is such that $r > \frac{\sqrt{2}d}{n_G}$, and for all $d_F$ dimensional faces $F$ of $[0,1]^d$ the reach $r_F$ of  $\partial f \cap F$ is such that $r_F > \frac{\sqrt{2}d_F}{n_G}$ then the Hausdorff distance between $\partial f \cap [0,1]^d$ and its resistar approximation $([B_f])^\star$ is in $\mathcal{O}(d n_G^{-2})$.
 %\label{th:accuracy1}
%\end{theorem} 
%\medbreak
%------------------------------------------------------
\subsection{Starting induction.}
%------------------------------------------------------
\label{sec:induc0}
%(by induction on the space dimensionality $d$).
Assume $d = 1$ and $r > \sqrt{2} \epsilon $. $M$ is a set of discrete points such that, for $b_M$ and $b'_M$  two distinct points of $M$, $\left\| b_M - b'_M\right\| >  \sqrt{2} \epsilon$. Therefore, in any edge $[v,v']$ of the grid, $v$ and $v'$ being two consecutive points such that $\left\|v-v'\right\| = \epsilon$, there is at most one point $b_M$ of $M$ in $[v,v']$ and for each point $b_M$ of $M$, there exists a single boundary point $b \in [v,v']$ with $\left\|b-b'_M\right\| \leq 2^{-q-1} \epsilon$, $q$ being the number of dichotomies performed to get $b$. Moreover, by construction, there is no boundary point in a segment $[v,v']$ such that $[v,v'] \cap M = \emptyset$. Therefore, choosing $q \geq -\log_2(\epsilon)$, ensures that theorem \ref{th:accuracy1} is true for $d = 1$.

Now, we assume (induction hypothesis) that the theorem is true in a compact of any dimensionality lower or equal to $d-1$ and we consider a manifold $M$ splitting a $X = [0,1]^d$ of dimensionality $d$ and its resistar approximation, both satisfying the conditions of theorem \ref{th:accuracy1}. In the next subsection, we bound the distance from the resistar approximation to $M$ and in the following subsection, we bound the distance from $M$ to the resistar approximation. In both cases, the neighbourhood of the boundary of $X$ is a specific case requiring the induction hypothesis.

%------------------------------------------------------
\subsection{Bounding the distance from the resistar approximation to $M$.} 
%------------------------------------------------------
\label{sec:induc1}

Let $C$ be a cube of the grid and $[B]^\star$ the resistar (c-resistar or K-resistar) approximation of $M$ in this cube, supposed non-void. 

\begin{figure}
 	\centering
 	\begin{tabular}{cc}
 		\fbox{\includegraphics[width=6.5 cm]{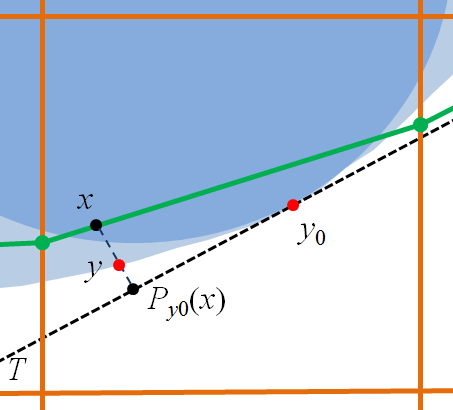}} & \fbox{	\includegraphics[width= 6.5 cm]{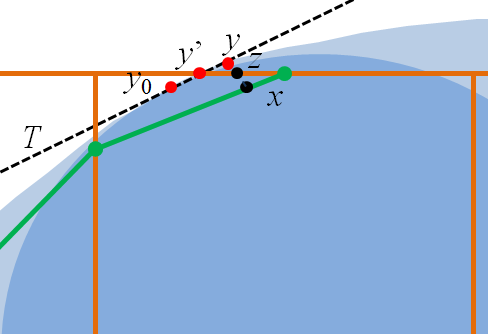} }\\
		(a) & (b)
 	\end{tabular}
 	\caption{Illustration of first part of proof of theorem \ref{th:accuracy1}. The resistar is represented in green and the limits of grid cubes in orange. $y = M \cap [z_-, z_+]$. Panel (a): $y \in X$, $\left\|x - y\right\|$ is bounded using $\left\|x - P_{y_0}(x)\right\|$ and $\left\|P_{y_0}(x) - y\right\|$. Panel (b): $y \notin X$. Point $y'$ on the boundary of $X$ is close to $z$, intersection of $[x,y]$ with the boundary of $X$, because of the induction hypothesis.}
 	\label{fig:accuracy1}
 \end{figure}

Since there are boundary points in $C$, $M$ cuts some edges of $C$ and $C \cap M \neq \emptyset$. Let $y_0 \in M \cap C$ and let $\mathcal{B}(c_+,r) $ and $\mathcal{B}(c_-,r) $ be the positive and negative balls tangent to $y_0$ as defined in the proof of lemma \ref{th:distTangent}. Applying lemma \ref{th:projTangent}, for all $x$ in $[B]^\star$, because the maximum distance between two points of $C$ is $\sqrt{d} \epsilon$, and $r > \sqrt{2} d \epsilon > \sqrt{d} \epsilon$ we have:
\begin{equation}
 \left\|x - P_{y_0}(x)\right\| \leq 2^{- q} \epsilon + \frac{d }{2r} \epsilon^2 + \mathcal{O}(\epsilon^{3}), 
\end{equation}	

where $P_{y_0}(x)$ is the orthogonal projection of $x$ on $T$, the hyperplane tangent to $M$ at $y_0$.
Moreover, $\left\| P_{y_0}(x) - y_0 \right\| \leq \sqrt{d}\epsilon $, because the orthogonal projection is contracting. Let $z_-$ and $z_+ $ be the projection of $x$ parallel to $n$, the normal vector to $T$, on respectively $\mathcal{B}(c_-,r)$ and $\mathcal{B}(c_+,r)$. Because of lemma \ref{th:distTangent}, we have (see figure \ref{fig:accuracy1} panel b):
\begin{equation}
 \left\|P_{y_0}(x) - z_-\right\| \leq  \frac{d}{2r} \epsilon^2 + \mathcal{O}(\epsilon^{3}) \mbox{  and  }
 \left\|P_{y_0}(x) - z_+\right\| \leq  \frac{d }{2r} \epsilon^2 + \mathcal{O}(\epsilon^{3}). 
\end{equation}	

There exists $y \in M \cap [z_-, z_+]$ because $M$ is a $d-1$-dimensional manifold located between $\mathcal{B}(c_+,r)$ and $\mathcal{B}(c_-,r) $, and we have:

\begin{equation}
 \left\|x - y\right\| \leq \left\|x - P_{y_0}(x)\right\| + \left\|P_{y_0}(x) - y\right\| \leq 2^{- q} \epsilon + \frac{d}{r} \epsilon^2 + \mathcal{O}(\epsilon^{3}),
\end{equation}	
because $P_{y_0}(x) = P_{y_0}(y)$ and $\left\|P_{y_0}(x) - y\right\| \leq \left\| P_{y_0}(x) - z_+ \right\|$ or  $\left\|P_{y_0}(x) - y\right\| \leq \left\| P_{y_0}(x) - z_- \right\|$.
Two cases can take place:
\begin{itemize}
	\item  $y \in X$, which is guaranteed when the cube $C$ is not at the boundary of $X$ (see figure \ref{fig:accuracy1}, panel (a)). Then if $q \geq - \log_2(\epsilon) $, $\left\|x - y\right\|  = \mathcal{O}(d \epsilon^2)$.
\item  $y \notin X$, which can happen when $C$ is at the boundary of $X$ (see figure \ref{fig:accuracy1}, panel (b)). Let $z = [x,y] \cap \partial X$ and let $F$ be the facet of $X$ such that $z \in F$. We have: $[x,z] \cap M = \emptyset$, hence $\mathcal{M}(z) \neq [B]^\star(z)$. Because of the induction hypothesis, there exists a point $y' \in (M \cap F)$ such that $\left\|y' - z\right\| = \mathcal{O}((d-1) \epsilon^{2})$.
 Since: 
\begin{equation}
\left\|x - z\right\| \leq \left\|x - y\right\| \leq 2^{- q} \epsilon + \frac{d }{r} \epsilon^2 + \mathcal{O}(\epsilon^{3}),
\end{equation}	
if $q \geq - \log_2(\epsilon) $, $\left\|x - y'\right\| \leq \left\|x - z\right\| + \left\|z - y'\right\| = \mathcal{O}(d \epsilon^2)$.

\end{itemize}

Therefore, in all cases, if $q \geq - \log_2(\epsilon) $, for all $x \in [B]^\star$, there exists $y \in M \cap X$ such that $\left\|x - y\right\| = \mathcal{O}(d \epsilon^2)$.

%------------------------------------------------------
\subsection{Bounding the distance from $M$ to its resistar approximation.} 
%----------------------------------------------------------------
\label{sec:induc2}

Consider a point $y_0 \in M \cap X$. Let $C$ be a cube of the grid such that $y_0 \in C$, let $T$ be the hyperplane tangent to $M$ at $y_0$, let $n$ be its normal vector and $\mathcal{B}(c_+,r) $ and $\mathcal{B}(c_-,r) $ be the positive and negative balls tangent to $y_0$. Let $\mathscr{C}$ be the cube of centre the centre of $C$, and of edge size $3 \epsilon$. The segment $[y_0,c_+]$ (resp. $[y_0,c_-]$) cuts the boundary of $\mathscr{C}$ at $z_+$ (resp. $z_-$), because $ r > \sqrt{2} d \epsilon > 2 \sqrt{d} \epsilon$ for $d \geq 2$. 

\begin{figure}
 	\centering
 		\fbox{\includegraphics[width=6.5 cm]{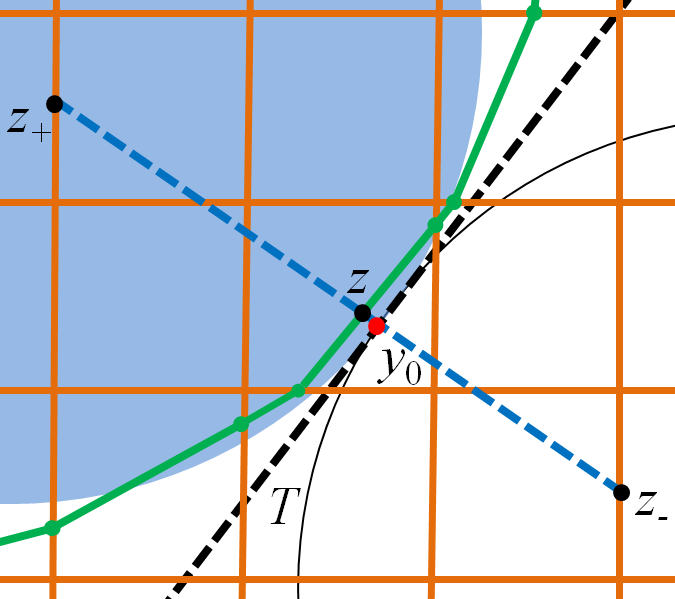}} 
 	\caption{Illustration of second part of the proof of theorem \ref{th:accuracy1}. The resistar is represented in green and the limits of grid cubes in orange. Cube $C$, in which $y_0$ is located, is not at the boundary of $X$. The $d-1$-dimensional cube face which includes $z_+$ is completely included in the positive tangent ball $\mathcal{B}(c_+,r)$ (represented in blue) and the $d-1$-dimensional cube face which includes $z_-$ is completely included in the negative tangent ball $\mathcal{B}(c_-,r)$ (whose boundary is represented in black).}
 	\label{fig:accuracy21}
 \end{figure} 

We first consider the case where there exists a cube $C'$ of the grid, adjacent to $C$ such that $z_+ \in C'$, then there exists $F$, facet of $C'$ such that $z_+ \in F$ (see Figure \ref{fig:accuracy21}).

We first show that $[B]^\star(z_+) = +1$ because the whole facet $F$ is included in $\mathcal{B}(c_+,r) $. Let $c'_+$ be the orthogonal projection of $c_+$ on the hyperplane $H_F$ defined by $F$. The intersection of $\partial \mathcal{B}(c_+,r)$ with $H_F$ is a sphere $\partial \mathcal{B}_F(c'_+, r')$ in $H_F$ of centre $c'_+$. Let $ \{ w \} = \vec{r}(c_+, z_+) \cap \partial \mathcal{B}_F(c'_+,r')$. $w$ is the point of $\partial \mathcal{B}(c_+,r) \cap H_F$ which is the closest to $z_+$. We have (see Figure \ref{fig:accuracy21}, panel (a)):

\begin{align}
	\left\|z_+ - w\right\| = r \frac{\sin \beta}{\cos \alpha}.
	\label{eq:projFace}
\end{align}

where $\alpha$ is the angle defined by $(c'_+, c_+, y_0)$ and $\beta$ the angle defined by $(y_0, c_+, w)$. The angle $\beta$ is minimum when the distance from $y_0$ to $F$ is $\epsilon$. Indeed, if $y_0$ gets closer to $H_F$, keeping the same angle between $H_F$ and $c_+ - y_0$, the radius $r'$ of the sphere $\partial \mathcal{B}_F(c'_+, r')$ decreases as well as the distance $\left\|w - z_+\right\|$ and so does $\beta$. Moreover, $\cos \alpha$ is maximum at 1. With these values, we have:

\begin{align}
	\sin \beta = \sqrt{1 - \left(\frac{r - \epsilon}{r}\right)^2}.
\end{align}

Therefore:
\begin{align}
	\left\|z_+ - w\right\| \geq \sqrt{ r^2 - (r - \epsilon)^2}.
	\label{eq:facet}
\end{align}

Then, expressing within equation \ref{eq:facet} that $\left\|z_+ - w\right\|$ is larger than $\sqrt{d-1} \epsilon$ the maximum distance between two points in the facet, requires:
\begin{align}
	r \geq \frac{d\epsilon}{2},
\end{align}

Therefore, as we assumed $r \geq \sqrt{2} d\epsilon$ this condition is satisfied, thus $\left\|z_+ - w\right\| \geq \sqrt{d-1} \epsilon$, implying that the whole facet $F$ is included in $\mathcal{B}(c_+,r)$ and $[B]^\star(z_+) = +1$.

Similarly, if there exists a cube $C'$ of the grid, adjacent to $C$ and such that $z_- \in C'$, then  $[B]^\star(z_-) = -1$.

Because of propositions \ref{prop:cseparationG} and \ref{prop:KseparationG}, the segment $[z_+,z_-]$ crosses the resistar (c-resistar or K-resistar) approximation $[B]^\star$ defined in $\mathscr{C}$ (with $B = B_M(\mathscr{C})$ or $B = B_M^K(\mathscr{C})$). Let $z \in [z_+,z_-] \cap [B]^\star$. Noticing that $P_{y_0}(z) = y_0$ and applying lemma \ref{th:projTangent}, we get: $\left\|y_0 - z\right\| = \mathcal{O}(d \epsilon^2)$.

\begin{figure}
 	\centering
 	\begin{tabular}{cc}
 		\fbox{\includegraphics[width=6.5 cm]{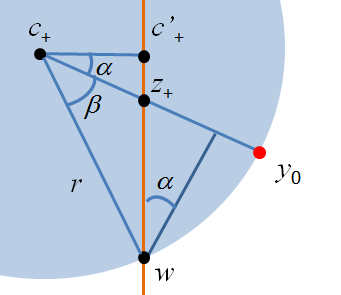}} & 	\fbox{\includegraphics[width= 6.5 cm]{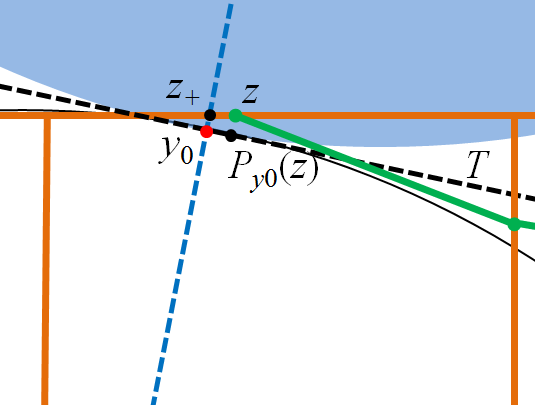}} \\
		(a) & (b)
 	\end{tabular}
 	\caption{Illustration of second part of the proof of theorem \ref{th:accuracy1}. Panel (a): Illustration of equation \ref{eq:projFace} estimating the distance $\left\|z_+ - w\right\|$. Panel (b): Cube $C$ is at the boundary of $X$ and $z_+$ is on the boundary of $X$. Because of the induction hypothesis, there exists a point $z$ in the resistar defined in the boundary which is close to $z_+$.}
 	\label{fig:accuracy2}
 \end{figure}

Now, we consider the case where the cube $C$ is on the boundary of $X$ and $[y_0,c_+]$ crosses the boundary of $X$ in $d-1$-dimensional facet $F$ of $C$, before crossing the boundary of $\mathscr{C}$. Let $z_+ = [y_0,c_+] \cap F$. Two cases can take place:
\begin{itemize}
	\item $[B]^\star (z_+) = +1$, then the same reasoning as previously applies;
	\item $[B]^\star (z_+) = -1$, is an error (because $z_+ \in \mathcal{B}(c_+,r)$) and because of the induction hypothesis, there exists a point $z$ in $[B_M(F)]^\star $ or in $[B_M^K(F)]$, such that $\left\|z_+ - z \right\| = \mathcal{O}((d-1)\epsilon^2)$. We have:
	\begin{align}	
		\left\|P_{y_0}(z) - P_{y_0}(z_+)\right\|  = \left\|P_{y_0}(z) - y_0\right\| = \mathcal{O}((d-1) \epsilon^2),
	\end{align}	
because the orthogonal projection is contracting, and:
\begin{align}	
 \left\|P_{y_0}(z) - z\right\| = \mathcal{O}(d \epsilon^2)
\end{align}
 because $\left\|z - y_0\right\| \leq \sqrt{d} \epsilon$ and applying lemma \ref{th:distTangent}. Therefore (see figure \ref{fig:accuracy2}, panel (b)):
\begin{align}	
\left\|y_0 - z\right\| \leq \left\|y_0 - P_{y_0}(z) \right\| + \left\|P_{y_0}(z) - z\right\| = \mathcal{O}(d \epsilon^2).
\end{align}
\end{itemize}

Of course, the same reasoning applies to the negative side $[y,c_-]$.
This concludes the proof of theorem \ref{th:accuracy1}.
$\square$ % \vspace{0.5 cm}

%==========================================================
\section{Examples and tests.}
%==========================================================

%----------------------------------------------------
\subsection{Visualisation of examples with spheres and radial based functions.}
%----------------------------------------------------

When the dimensionality $d$ is higher than 3, the surface cannot be directly represented, but it is possible to visualise its intersection with $d - 3$	hyperplanes. Algorithm \ref{Algo:visu} sketches the method. It is based on building polytopes whose vertices are the intersections of edges of another polytope with a hyperplane. Starting with a simplex of the resistar, a polytope of one dimension less is computed in this way successively with each of the hyperplanes. If the intersection of the resistar simplex with the intersection of the $d-3$ hyperplanes is not empty, the result is a 2D polygon in the 3D intersection of the $d - 3$ hyperplanes. These polygons define together a 2D surface in this 3D space.

\begin{algorithm}%[T]
%\dontprintsemicolon
\KwIn{$[B]^\star$ resistar approximation, $H_i$, $i \in \{1,..., d-3\}$ hyperplanes cutting $X$.}
	$L \leftarrow \emptyset$;\\
	\For{$[S] \mbox{ simplex of }[B]^\star, [S] \cap H_i \neq 0$, $i \in \{1,..., d-3\}$}{
	$P \leftarrow S$;\\
	\For{$i \in \{1,..., d-3\}$}
	{$\begin{aligned}
	P \leftarrow \bigcup_{[v,v'] \mbox{ edge of [P]}} [v,v'] \cap H_i
	\end{aligned}$}
	\If{ $[P] \neq \emptyset$}{$L \leftarrow L \cup [P]$}
	}
\Return $L$;
\label{Algo:visu}
\caption{Intersection of a resistar approximation with $d-3$ hyperplanes cutting $X$.}
\end{algorithm} 

Even though it is possible to focus on a small percentage of all the simplices of the resistar approximation that have chances to intersect with the intersection of all the hyperplanes, this small percentage may still correspond to a high number of simplices in some cases. This can occur easily when the dimensionality of the space is higher than 6, especially with K-resistars.   

\begin{figure}
 	\centering
 	\begin{tabular}{cc}
 			\includegraphics[width = 7.0 cm]{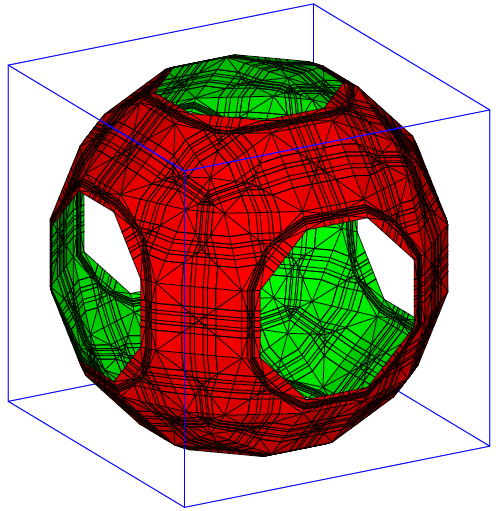} \hspace{0.5 cm} & \hspace{0.5 cm}
			 	\includegraphics[width= 7.0 cm]{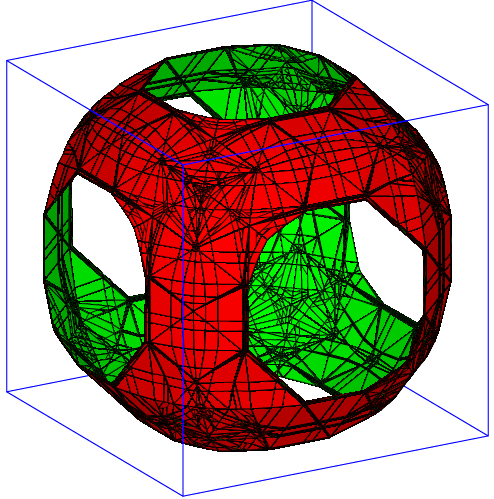} \\
				(a) & (b) 
					%\includegraphics[width= 7.5 cm]{figuresFormes/2SpheresClose7D-s.pdf} &
			 	%\includegraphics[width= 7.5 cm]{figuresFormes/2SpheresClose7D-m.pdf} \\
				%(c) & (d) 
 	\end{tabular}
 	\caption{Examples of resistar approximations with $n_G = 5$ points by axis of the grid. In panel (a), a c-resistar approximation of a sphere in 6 dimensions. In total the resistar includes 11,520 boundary points and $7.3$ million simplices. The represented 3D surface is the intersection of the resistar surface with hyperplanes $x_4 = 0.46$, $x_5 = 0.49$ and $x_6 = 0.48$. It includes 6,144 polygons. In panel (b), a K-resistar approximation of a sphere in 5 dimensions. In total the resistar includes 27,102 boundary points and $12.9$ million simplices. The represented 3D surface is the intersection of the resistar surface with hyperplanes of equations $x_4 = 0.46$ and $x_5 = 0.49$. It includes 585,150 polygons.}
 	\label{fig:Examples1}
 \end{figure}

Figure \ref{fig:Examples1} shows examples of approximations of a sphere in 5D with K-resistars (panel (b)) and in 6D with c-resistars (panel (a)), for $n_G = 5$ points on each axis of the grid. Note that the K-resistar includes a larger number of boundary points and even larger number of simplices, although it is in 5 dimensions while the c-resistar is in 6 dimensions. In 6 dimensions, the K-resistar approximating the same sphere for $n_G = 5$ involves 199,322 boundary points and $1.1$ billion simplices. 
	
 	 %\begin{figure}
 	%\centering
 	%\begin{tabular}{c}
 		%
			%\includegraphics[width= 12 cm]{figuresFormes/2ClassesRadialD3N48pp300pn300s003.pdf}
 	%\end{tabular}
 	%\caption{Example of m-resistar surface in 3D from $M$ defined by equation \ref{eq:radial}, with 300 positive and 300 negative points drawn at random, parameter $\sigma = 0.03$ and for a grid with 48 points by dimension ($n_G = 48$). The resistar surface includes 214,358 simplices in total.}
 	%\label{fig:Examples2a}
 %\end{figure} 

 	 \begin{figure}
 	\centering
 	\begin{tabular}{cc}
 			 	\includegraphics[width= 7.5 cm]{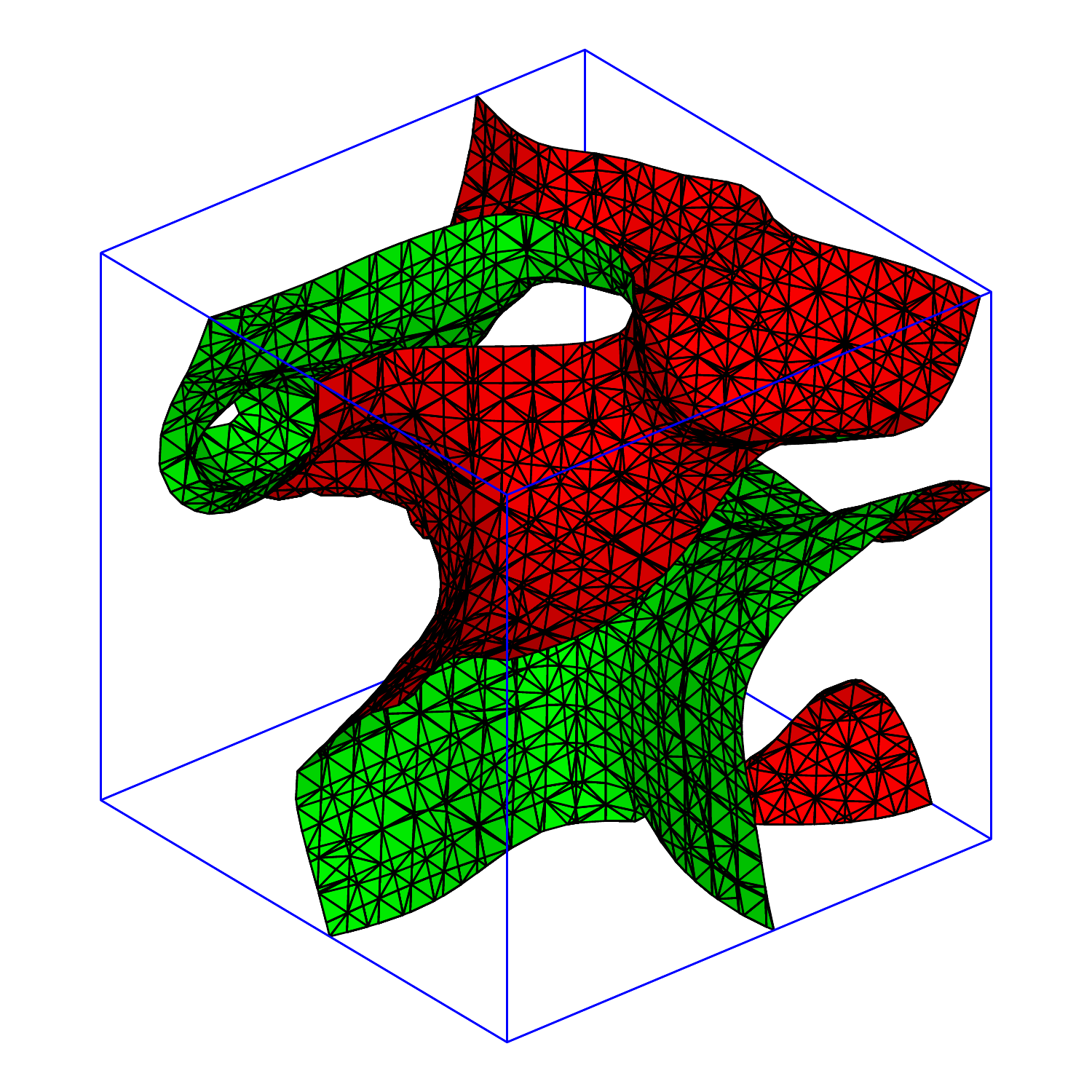} & \includegraphics[width= 7.5 cm]{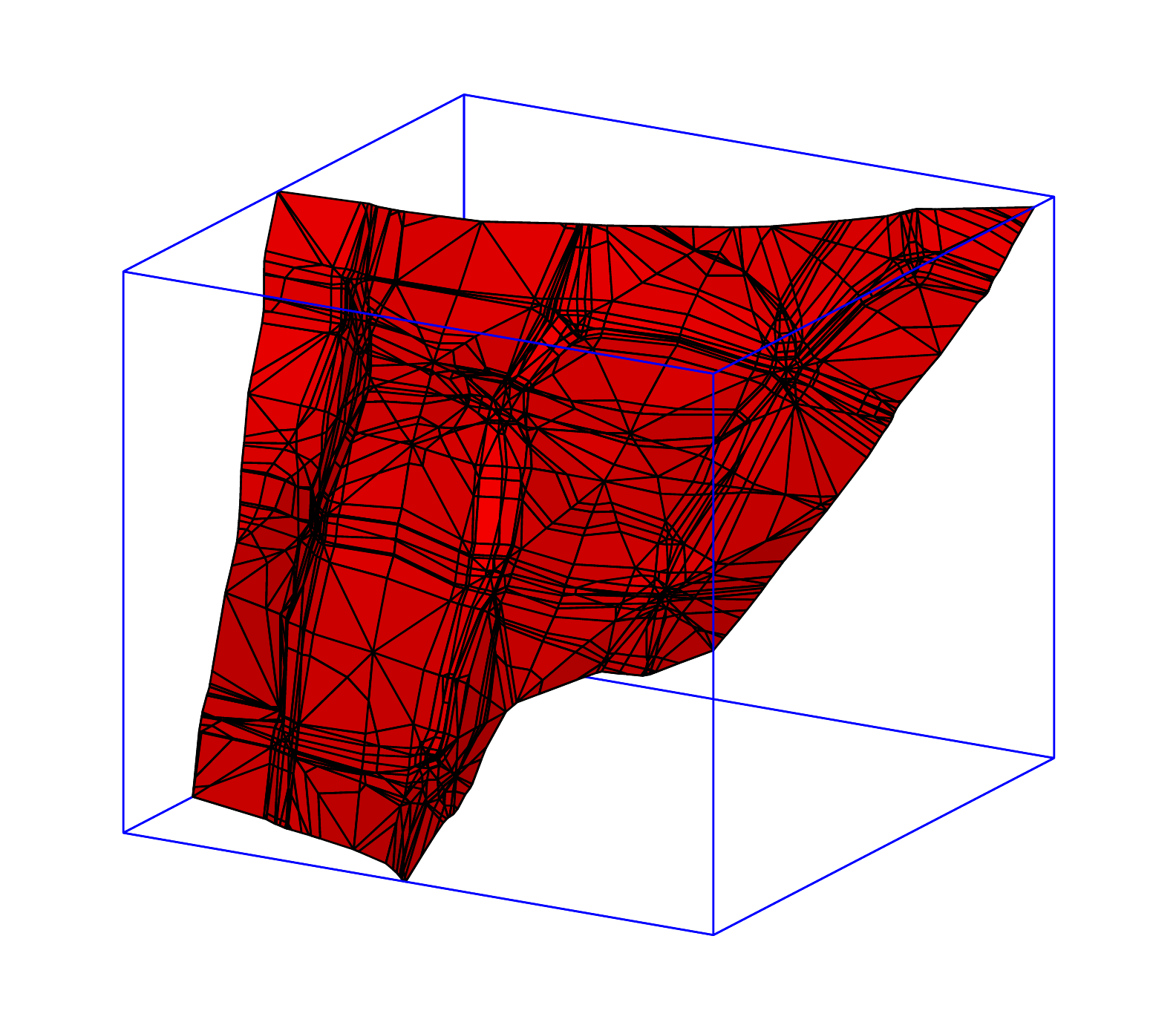}\\
				(a) & (b)
 	\end{tabular}
 	\caption{Approximation of surfaces derived from radial based functions (see  equation \ref{eq:radial}). Panel (a): K-resistar approximation in 3 D, on a grid with 16 points by dimension ($n_G = 16$). The manifold $M$ to approximate is derived from a radial based function with the size of $E_p$ and $E_n$ equal to 20 and parameter $\sigma = 0.2$. The K-resistar approximation includes 3,517 boundary points and 28,950 simplices. The estimated error percentage is: $0.3 \%$. Panel (b): c-resistar approximation in 7D, on a grid with 4 points by dimension ($n_G = 4$). The manifold $M$ to approximate is derived from a radial based function with the size of $E_p$ and $E_n$ equal to 10 and parameter $\sigma = 0.4$. The resistar approximation includes 11,285 boundary points and 21.8 million simplices in total. The estimated error percentage is: $2.07 \%$. The represented 3D surface is the intersection of the resistar surface with the hyperplanes of equations $x_4 = 0.47$, $x_5 = 0.48$, $x_6 = 0.49$ and $x_7 = 0.55$.}
 	\label{fig:Examples2}
 \end{figure}

  	 %\begin{figure}
 	%\centering
 	%\begin{tabular}{cc}
 		%\includegraphics[width=8 cm]{figuresFormes/sphereD6N5R048.pdf} 
 		 	%\includegraphics[width= 8 cm]{figuresFormes/sphereD6N5R061.pdf}
 	%\end{tabular}
 	%\caption{Isosurface approximations with simplex stars. 6D spheres with $n_c = 5$, the 3D representation is done by cutting the shape with hyperplanes of equations $x_1 = 0.49$, $x_2 = 0.48$, $x_3 = 0.47$. On the left the radius is 0.48 and the total number of simplexes is 8,488,287. On the right, for a radius of the sphere of 0.61, the total number of simplexes is 18,554,802.}
 	%\label{fig:Examples3}
 %\end{figure} 
 
In order to test more systematically our approach, we derived classifications from a radial-based function defined on two sets of randomly chosen points in the space $E_p = \left\{p_1, p_2,.., p_l\right\}$ and $E_n = \left\{n_1, n_2,.., n_m\right\}$, as follows:
\begin{equation}
\mathcal{M}(x) = \mbox{sign}\left(\sum_{i = 1}^l \phi \left(\frac{p_i - x}{\sigma}\right) - \sum_{i = 1}^m \phi \left(\frac{n_i - x}{\sigma}\right)\right),
\label{eq:radial}
\end{equation}
where $\sigma$ is a positive number and function $\phi$ is defined by:
\begin{equation}
\phi(x) = \frac{100}{1 + x^2}.
\end{equation}

When varying the number of points in $E_p$ and $E_n$ and the value of $\sigma$, the resulting surface is more or less complicated and smooth. In our tests, we use two settings that are illustrated by Figure \ref{fig:Examples2}.

%----------------------------------------------------
\subsection{Testing the classification error when the grid size and the space dimensionality vary.}
%----------------------------------------------------
Generally, the accuracy of classification methods is measured by the misclassification rate of points randomly drawn in $[0,1]^d$. However, this measure requires drawing a very large number of test points when the dimensionality $d$ of the space increases. In order to limit the number of tests, we generated the test points only in the cubes which include boundary points. This gives more chances to get misclassified points. Taking 100 test points in each cube containing boundary points, we used 50 million test points for the resistar approximation with the highest number of cubes in the following tests (in dimensionality 4 with $n_G = 48$).

\begin{figure}%[H]
 	\centering
			\fbox{\includegraphics[width= 9 cm]{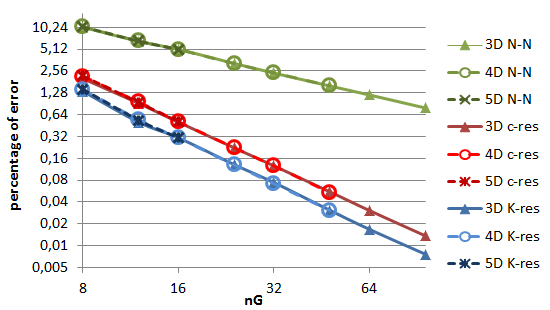}} %&
		 	%\includegraphics[height= 6 cm]{figures/nbTestPointsNg.png} \\
			%(a) & (b)
 	\caption{Classification error percentage for K-resistars, c-resistars and nearest vertex when $n_G$ (number of points by axis of the grid) varies and for different values of dimensionality $d$. The manifold $M$ to approximate is derived from a radial based function with the size of $E_p$ and $E_n$ equal to 20 and $\sigma = 0.2$. }
 	\label{fig:classifNg}
 \end{figure} 	

Figure \ref{fig:classifNg} shows that the classification error decreases like $n_G^{-2}$ for both K-resistars and c-resistars (estimated slopes in the log-log graph: -2.06, for K-resistars, -2.01 for c-resistars, with $R^2 = 0.99$ for both) in accordance with theorem \ref{th:accuracy1}, whereas the classification error of nearest vertex decreases like $n_G^{-1}$ (estimated slope in the log-log graph: -1.02, with $R^2 = 0.99$). It appears also that the error of K-resistars is significantly lower than the one of c-resistars, which is expected because K-resistars are based on a larger number of boundary points located on the edges of the Kuhn simplices. Note that the values of errors do not change significantly when the dimensionality changes.

\begin{figure}%[H]
 	\centering
 	\begin{tabular}{c}		
			\fbox{\includegraphics[width= 9 cm]{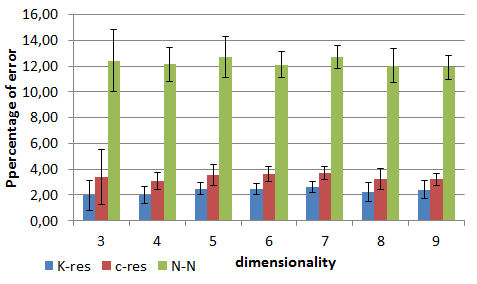}} 
 	\end{tabular}
 	\caption{Classification error percentages of nearest vertex, c-resistar and K-resistars approximating radial based classification surface with the size of $E_p$ and $E_n$ equal to 10, $\sigma = 0.4$ and $n_G = 4$, in space of dimensionality varying from 3 to 9. The test points are uniformly drawn in the cubes which include boundary points. The error bars correspond to the standard deviation over 10 different functions defined with new points $E_p$ and $E_n$. }
 	\label{fig:classifNNMS}
 \end{figure}

This is confirmed by Figure \ref{fig:classifNNMS} which shows the classification error for nearest vertex, c-resistar and K-resistars approximating radial based classification functions with the same parameters ($E_p$ and $E_n$ equal to 10, $\sigma = 0.4$) and the same number of points by axis of the grid $n_G = 4$, in dimensionality varying from 3 to 9. For each dimensionality, the tests are repeated for 10 radial-based classification functions with points $E_p$ and $E_n$ drawn at random in $[0,1]^d$. The error percentage is computed by classifying 100 points uniformly drawn in each cube which includes boundary points. We observe that the error percentages do not vary significantly with the dimensionality whereas the bound on the Hausdorff distance should vary linearly with $d$. 

\begin{figure}%[H]
 	\centering
 	\begin{tabular}{cc}		
			\fbox{\includegraphics[width= 8 cm]{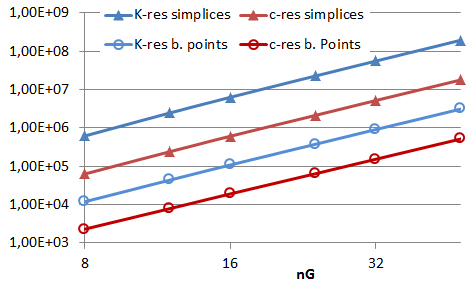}} &  
			\fbox{\includegraphics[width= 8 cm]{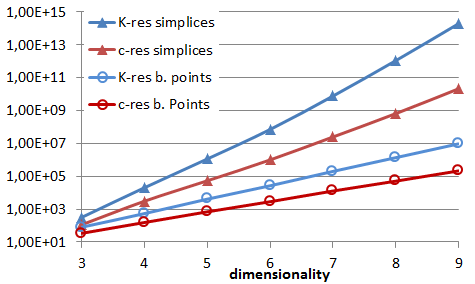}} \\
			(a) & (b)
 	\end{tabular}
 	\caption{Numbers of boundary points and of simplices in c-resistars and K-resistars. Panel (a): In space of 4 dimensions, with $n_G$ varying from 8 to 48, for a radial based function with the size of $E_p$ and $E_n$ equal to 20, $\sigma = 0.2$. Panel (b): For $n_G = 4$ in space of dimensionality varying from 3 to 9, for a radial based function with $E_p$ and $E_n$ equal to 10, $\sigma = 0.4$. }
 	\label{fig:bpsSimplices}
 \end{figure}

Figure \ref{fig:bpsSimplices} shows the number of boundary points and simplices when the number of points of the grid or the dimensionality of the space vary. These graphs confirm the very rapid growth of the number of boundary points and of simplices, particularly in K-resistars. In 9 dimensions, for $n_G =4$, the number of boundary points of the K-resistar surface is around 10 million, and the number of simplices is around $10^{14}$ (see panel (b)). There are about 100 times less boundary points and 10,000 times less simplices in the c-resistar surface. 
The estimation of the slope of the logarithm of the number of simplices as a function of the logarithm of $n_G$ for the tests in dimensionality 3, 4 and 5 presented on Figure \ref{fig:classifNg} (see the case of dimensionality 4 on figure \ref{fig:bpsSimplices} panel (a)) is reported in table \ref{table:slopes}. For c-resistars, the number of boundary points appears thus to grow like $\mathcal{O}(n_G^{d-1})$. For K-resistars the slope is a bit higher than $d-1$ and the difference increases with the dimensionality. This is due to the number of edges in all the K-simplices in a cube which increases much more rapidly than the number of edges of the cube.

\begin{table}[h]
\centering
\begin{tabular}{|c|c|c|c|}
\hline
Dimensionality $d$ & 3 & 4 & 5\\
\hline
c-resistars & 2.05 & 3.02 & 4.05 \\
K-resistars & 2.05 & 3.10 & 4.23\\
\hline
\end{tabular}
\label{table:slopes}
\caption{Estimated slope of the logarithm of the number of simplices in the resistar approximation as a function of the logarithm of $n_G$ the number of points by axis of the grid.}
\end{table}

%==========================================================
\section{Discussion - conclusion}
%==========================================================

This paper shows examples of simplex based approximation and classification in 9 dimensions, which has never been done with marching cube or Delaunay triangulation. Indeed, the resistar classification is achieved through a few projections on facets and faces of a cube while the other methods would require to test the position of the point to classify with respect to a large number of simplices. This advantage of resistars starts in low dimensionality and becomes decisive in higher dimensionality as the number of simplices increases exponentially. 

The classification methods used in the algorithms of viability kernel approximation, such as nearest vertex, SVM or k-d trees, are based only on the classification of the vertices of a regular grid, hence their error cannot decrease more than $\mathcal{O}(n_G^{-1})$ which is the intrinsic error in the learning sample. Resistar approximation does better because it is based on the boundary points which can be at a precision of $\mathcal{O}(n_G^{-2})$ with an adequate number of dichotomies. Some other methods, for instance decision trees \cite{Chaturvedi2013,Dasgupta2015} could possibly be modified to learn efficiently from boundary points, but such a modification does not seem immediate. This specificity of resistar approximations allows them to ensure, under arguably reasonable conditions, that their Hausdorff distance to the manifold to approximate decreases like $\mathcal{O}(n_G^{-2})$. This is a very significant advantage over current methods. Indeed, the resistar classification from the a grid of $n_G^{d}$ points has the same accuracy as a standard classification based on a grid of $(n_G^2)^d$ points. 

Computing the boundary points of resistars requires first classifying by $\mathcal{M}$ the $n_G^{d}$ grid points. Then, for c-resistars which generate $\mathcal{O}(n_G^{d-1})$ boundary points, there are in total $\mathcal{O} (q. n_G^{d-1})$ point classifications by $\mathcal{M}$, because of the $q$ successive dichotomies necessary to compute each boundary point. If $q = \log_2 (n_G)$ as in Theorem \ref{th:accuracy1}, the number of point classifications by $\mathcal{M}$ required by c-resistars is $n_G^{d} + log(n_G). n_G^{d-1}$, which is very significantly lower than the  $n_G^{2d}$ grid vertex classifications required by standard methods to get the same accuracy. For K-resistars, the number of boundary points increases also approximately like $\mathcal{O} (n_G^{d-1})$ in low dimensionality, but this number may be closer to $\mathcal{O}(n_G^{d})$ in dimensionality 8 or 9. Overall, the number of classifications by $\mathcal{M}$ remains still very significantly lower than the one required by usual methods to get the same accuracy. Considering the requirements in memory space, the advantage of resistars is very strong over the nearest vertex classification which needs to store the whole grid of $n_G^{2d}$ points (or with some optimisation $\mathcal{O}(n_G^{2d-1})$ points) whereas the c-resistars need to store $\mathcal{O}(n_G^{d-1})$ boundary points and K-resistars at worst $\mathcal{O}(n_G^{d})$, to get the same accuracy.

The two variants of resistars have different strengths and weaknesses. The K-resistars approximations have the major advantage to be manifolds and their error rate is lower than the one of the c-resistars for a given grid size. However, the c-resistars are significantly lighter, especially when the dimensionality increases. In some cases, it is possible to use them while the K-resistars are too heavy. 

For both of them, the accuracy in $\mathcal{O}(n_G^{-2})$ requires the manifold to approximate to be smooth, which is not always the case in viability problems. This manifold is indeed often the boundary of the intersection of several smooth manifolds, hence with a reach equal to zero. A challenging future work is to define new types of resistars approximating these intersections with an accuracy of $\mathcal{O}(n_G^{-2})$ and with an efficient classification algorithm.

%The resistar approach could also be interesting for generating meshes in spaces of dimensionality higher than 4. The number of simplices generated in 5 or 6 dimensions is of the order of $10^6$ to $10^9$, and this seems manageable with current high performance computing, especially because the simplices of the resistars can be computed rapidly when needed, limiting the storage to the boundary points in non-empty cubes. However, an important problem is that m-resistars may include multiple singularities in dimensionality higher than 3, if the approximated region on the cube boundary is not homomorphic to a sphere (for instance when it is homomorphic to a torus). Future work would be necessary in order to detect these configurations and then to modify adequately the resistar definition. 

%=======================================================================================
\section{Acknowledgment} 
%=======================================================================================
I am grateful to Sophie Martin and Isabelle Alvarez for their helpful comments and suggestions on earlier versions of the paper.

\section{References} 

\bibliographystyle{elsarticle-num}
\bibliography{starMC}

\end{document}